\documentclass[%
 reprint,
superscriptaddress,
 amsmath,amssymb,
 aps,
]{revtex4-2}

\usepackage{graphicx}
\usepackage{dcolumn}
\usepackage{bm}
\usepackage{subcaption}
\usepackage[version=4]{mhchem}
\usepackage{siunitx}

\begin{document}

\preprint{APS/123-QED}

\title{Magneto-optic Response of the Metallic Antiferromagnet \ce{Fe2As} to Ultrafast Temperature Excursions}

\author{Kexin Yang}
\affiliation{Department of Physics, University of Illinois at Urbana-Champaign, Urbana, Illinois, 61801}
\affiliation{Materials Research Laboratory, University of Illinois at Urbana-Champaign, Urbana, Illinois, 61801}
\author{Kisung Kang}
\affiliation{Materials Research Laboratory, University of Illinois at Urbana-Champaign, Urbana, Illinois, 61801}
\affiliation{Materials Science and Engineering, University of Illinois at Urbana-Champaign, Urbana, Illinois, 61801}
\author{Zhu Diao}
\affiliation{Materials Research Laboratory, University of Illinois at Urbana-Champaign, Urbana, Illinois, 61801}
\affiliation{Department of Physics, Stockholm University, SE-106 91 Stockholm, Sweden}
\affiliation{School of Information Technology, Halmstad University, P.O. Box 823, SE-301 18 Halmstad, Sweden}
\author{Arun Ramanathan}
\author{Manohar H. Karigerasi}
\author{Daniel P. Shoemaker}
\author{Andr\'e Schleife}
\thanks{schleife@illinois.edu}
\author{David G. Cahill}
\thanks{d-cahill@illinois.edu}
\affiliation{Materials Research Laboratory, University of Illinois at Urbana-Champaign}
\affiliation{Materials Science and Engineering, University of Illinois at Urbana-Champaign}

\date{\today}

\begin{abstract}
The linear magneto-optical Kerr effect (MOKE) is often used to probe magnetism of ferromagnetic materials, but MOKE cannot be applied to collinear antiferromagnets (AFs) due to the cancellation of sub-lattice magnetization. Magneto-optical constants that are quadratic in magnetization, however, provide an approach for studying AFs on picosecond time scales. Here, we combine transient measurements of optical reflectivity and birefringence to study the linear optical response of \ce{Fe2As} to small ultrafast temperature excursions. We performed temperature dependent pump-probe measurements on crystallographically isotropic (001) and anisotropic (010) faces of \ce{Fe2As} bulk crystals. We find the largest optical signals arise from changes in the index of refraction along the $z$-axis, i.e.\ perpendicular to the N\'eel vector. Both real and imaginary parts of the time-resolved optical birefringence rotation signal approximately follow the temperature dependence of the magnetic heat capacity, as expected if the changes in dielectric function are dominated by contributions of exchange interactions to the dielectric function. We conclude that under our experimental conditions, changes in the exchange interaction contribute more strongly to the temperature dependence of the magneto-optic constants than the Voigt effect.
\end{abstract}

\maketitle
\section{Introduction}

Antiferromagnetic materials are under intense investigation as a new generation of spintronic materials because of their robustness to external magnetic fields and ultrafast dynamics, as it manifests itself, for instance, in a higher resonance frequency, compared to ferromagnets \cite{duine2011spintronics,sinova2012new,SatohCr2O3,SatohNiO,PisarevReview}.
Characterization of the structure and dynamics of the magnetic order parameter is essential for spintronics research but is difficult to achieve in antiferromagnets (AFs). Magneto-optical effects are often a valuable tool for probing magnetic order; for example, much of what is known about the dynamics of ferromagnetic and ferrimagnetic materials comes from studies that make use of the linear magneto-optical Kerr effect (MOKE)  \cite{Bigot_Ni,FiMMOKE}. Linear MOKE is also an essential tool for imaging the structure of magnetic domains \cite{Bigot_Ni,Kerrmicroscopy}. For typical AFs, however, linear MOKE is absent. Application of linear MOKE in the study of AFs is mostly limited to AFs with weak ferromagnetism due to canted magnetic moments, e.g., in orthoferrites \cite{cantedAFM}. More recently, relatively large linear magneto-optic effects were observed in the \emph{non-collinear} AF \ce{Mn3Sn}\cite{Mn3Sn,Kimelreview}.

The structure and dynamics of the order parameter of AFs is typically probed using interactions that are quadratic in the magnetization.  For example, anisotropic magnetoresistance (AMR) depends on contributions to electronic relaxation times that are quadratic in magnetization; AMR is sensitive to the domain structure of antiferromagnetic materials\cite{MuirCr}.  More recently, AMR was used to read the spin configuration of antiferromagnetic CuMnAs\cite{jungwirth2016antiferromagnetic} and \ce{Mn2Au}\cite{Mn2Au}. At x-ray wavelengths,  magnetic linear dichroism (XMLD) probes the anisotropy of charge distributions that are quadratic in the magnetization \cite{XMLD_APL}.

Magnetic birefringence refers to anisotropies in the optical frequency dielectric function that are generated by terms that are second order in the magnetization.  Since the dielectric function and the second order terms of magnetization are both second rank tensors, the quadratic magneto-optic coefficients are a fourth rank tensor. Magnetic birefringence has been widely used in studies in optically transparent  AFs \cite{ferre1984linear} and ferromagnetic garnets.  

In 2017, Saidl {\it et~al.}\cite{CuMnAs} reported their studies of the time-resolved magneto-optic response of AF CuMnAs to a large temperature excursion, $\Delta T\sim$100~K. CuMnAs films were grown epitaxially on GaP(001) substrates with the $z$-axis, which is the hard magnetic axis of \ce{CuMnAs}, parallel to the surface normal. The magnetic structure of tetragonal CuMnAs has two degenerate magnetic domains with perpendicular N\'eel vectors in the $x$-$y$ plane.  For a 10 nm thick CuMnAs layer, the authors observed a rotation of the polarization of the optical probe beam that is consistent with an optical response that is quadratic in  magnetization, $\Delta\theta\propto\sin{2\alpha}$, where $\alpha$ is the angle between the N\'eel vector and the light polarization.

In our work, we studied transient changes in the optical frequency dielectric function of the metallic AF \ce{Fe2As}, produced by a small temperature excursion, $\Delta T\sim$3~K. We acquire data for changes in birefringence and reflectivity using techniques that we refer to as time-domain thermo-birefringence (TDTB) and time-domain thermo-reflectance (TDTR). TDTB and TDTR signals are acquired using a pump-probe apparatus based on a high repetition rate Ti:sapphire laser oscillator operating at a wavelength near 785 nm.  

\ce{Fe2As} crystallizes in the \ce{Cu2Sb} tetragonal crystal structure with easy-plane magnetization as shown in Fig.~\ref{fig:structure} \cite{Fe2Asstructure}. 
Based on the corresponding magnetic symmetry (mmm1' magnetic point group), the N\'eel vector of \ce{Fe2As} has two degenerate orientations in the $\left<100\right>$ and $\left<010\right>$ directions. Hence, on length scales large compared to the domain size, the dielectric function in the $x$-$y$ plane is isotropic. We indeed do not observe a significant TDTB signal for the (001) surface of \ce{Fe2As}; however, on the crystallographically anisotropic (010) surface of the tetragonal crystal, we observed a strong TDTB signal for light polarized at an angle of 45$^\circ$ between the $x$ and $z$ axis of the crystal. We gain complementary insight by measuring the TDTR signals for light polarized along the $x$ and $z$ axes.

Typically, the dominant contribution to the magnetic birefringence of ferromagnetic materials is the Voigt effect, where only the components of the dielectric tensor \emph{parallel} to the magnetization are affected \cite{LeGall,tesavrova3Dtrajectory}.
However, as we discuss below, our data implies $\Delta \varepsilon_{\perp}\gg\Delta\varepsilon_{\parallel}$ and the dependence of the TDTB signal on the sample temperature closely resembles the temperature dependence of the magnetic heat capacity, supporting a direct connection between exchange energy and dielectric function.
Hence, if we assume that magnetization is the only contribution for a  temperature dependence of the dielectric function, we conclude that, in \ce{Fe2As}, the quadratic magnetization term contributes most strongly to the dielectric function \emph{perpendicular} to the N\'eel vector, implying that the Voigt effect is not the major contribution to the TDTB signal of \ce{Fe2As}. 

Our experiments also provide insight into the ultrafast magnetization dynamics of \ce{Fe2As}.  Furthermore, by comparing changes in the magneto-optical response at short and long time-scales, we evaluate the importance of magnetostriction to magnetic birefringence in this material. 

\section{Experiments}
\subsection{Experimental Details}

Single crystals of \ce{Fe2As} were synthesized from the melt. Stoichiometric amounts of elemental Fe and As (99.8\% and 99.999\%, Alfa Aesar) were ground inside an argon filled glove box in an agate mortar and pestle. The powder mixture was loaded in a 6 mm-diameter fused silica tube and sealed under vacuum. The tube was heated to 700 $^{\circ}$C and held for 24~h, then 1000 $^{\circ}$C for 2~h, with 5 $^{\circ}$C/min ramp rate. The tube was cooled to 900 $^{\circ}$C in 20~h, then cooled at 5 $^{\circ}$C/min to obtain shiny gray crystals of \ce{Fe2As}. The phase purity of the sample was confirmed using powder X-ray diffraction on a Bruker D8 diffractometer with Mo \textit{K$\alpha$} source and LYNXEYE XE detector in the transmission geometry. Rietveld refinements were performed using TOPAS 5. The lattice constants at room temperature are $a=3.63$~\AA, $c=5.98$~\AA.

Before optical measurements, the chunk \ce{Fe2As} sample was polished along the (001) and (010) orientation with an Allied Multiprep automatic polisher with diamond lapping films down to 0.3 \si{\mu m}. The orientation was observed via X-ray diffraction pole figures. The miscut of the surfaces is within 10$^{\circ}$.  After polishing, the sample was ion milled for 5 min by an ion miller with Ar source of 250 V beam voltage and 60 mA beam current. 

The TDTB and TDTR measurements were done with a pump-probe system that employs a Ti:sapphire laser with a 80 MHz repetition rate and 783 nm cener wavelength. The spectral linewidth is 10 nm and the full-width-half-maximum of the pump-probe correlation is 1.1 ps. The pump beam is modulated at 10.8 MHz while the probe beam is modulated at 200 Hz. A half-wave plate was placed on the probe beam path to rotate the polarization of probe beam. For TDTB experiment, the light polarization or the ellipticity change was captured by a balanced photodetector, while in TDTR experiment, the transient reflection was measured by a Si photodetector. Both TDTB and TDTR signals were double modulated to reduce the background created by diffusively scattered pump and coherent pickup of electronics. In these experiments, the $1/e$ laser spot size of both pump and probe is 5.5~\si{\mu m}, and the laser fluence of pump is 0.22~\si{J/m^2}, which created a steady-state heating of 13 K on the sample surface. The zero time delay was determined with a GaP two-photon detector.  For temperature-dependent measurements, the sample was mounted on a heating stage in a vacuum of $10^{-3}$ Torr. 

We performed first-principles calculations using density functional theory (DFT) as implemented in the Vienna \emph{Ab-Initio} Simulation Package \cite{Kresse:1996,Kresse:1999,Gajdos:2006} (VASP). The generalized-gradient approximation (GGA) formulated by Perdew, Burke, and Ernzerhof \cite{Perdew:1997} (PBE) is used to describe exchange and correlation. The projector-augmented wave \cite{Blochl:1994} (PAW) scheme is used to describe the electron-ion interaction. To sample the Brillouin zone a $15\,\times\,15\,\times\,5$ Monkhorst-Pack \cite{Monkhorst:1976} (MP) $\mathbf{k}$-point grid is used and the Kohn-Sham states are expanded into plane waves up to a cutoff energy of 600 eV. Total energies are converged to self consistence within $10^{-6}$ eV. 
Noncollinear magnetism and spin-orbit coupling are included and the magnetic unit cell of Fe$_2$As is used to compute relaxed atomic geometries, electronic structure, and optical properties. Phonon dispersion is computed using finite displacement method as implemented in VASP and extracted using the phonopy package \cite{Phonopy:2015}. After convergence test, a $3\,\times\,3\,\times\,2$ supercell and $4\,\times\,4\,\times\,4$ MP $\mathbf{k}$-point grid is used. For phonon calculation, noncollinear magnetism and spin-orbit coupling is included.

\begin{figure}[htb]
\centering
\begin{subfigure}{0.2\textwidth}
		\centering
		\includegraphics[width=\textwidth]{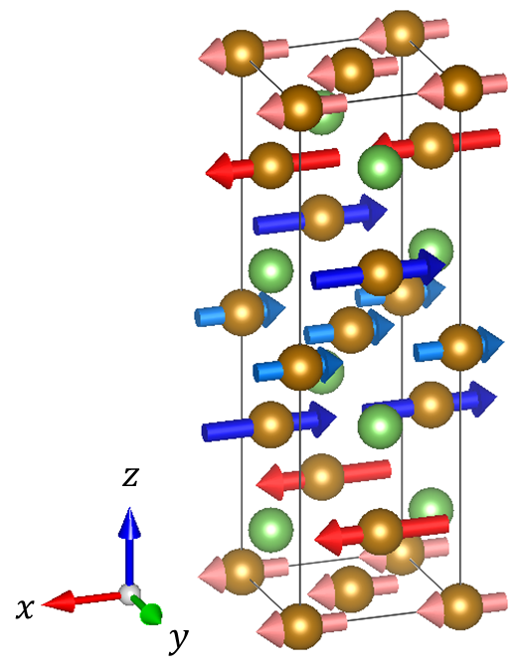}
		\caption[]
		{{\small }}   
		\label{fig:structure}
\end{subfigure}
\\
\begin{subfigure}{0.15\textwidth}  
		\centering 
		\includegraphics[width=\textwidth]{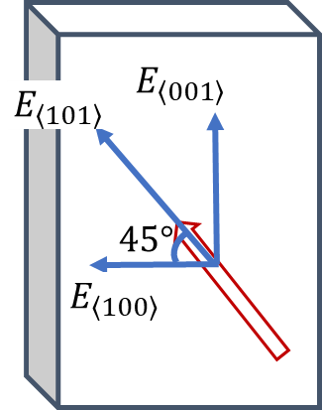}
		\caption[]
		{{\small }} 
		\label{fig:expr_diagram}
\end{subfigure}
\label{fig:structure_setup}
\caption[]
{
(a) Tetragonal magnetic unit cell  of \ce{Fe2As}. Arsenic atoms are depicted as green spheres; Fe as brown spheres. Arrows denote the local magnetic moment of the Fe atoms. Fe atoms labelled with blue and pink arrows are crystallographically equivalent. The Cartesian coordinates $x,y,z$ are aligned along the crystallographic $a,b,c$ axes. (b) Experimental geometry for the time-domain thermo-birefringence (TDTB) and time-domain thermoreflectance (TDTR) experiments with the probe beam normal to the (010) face of the \ce{Fe2As} crystal. In TDTR measurements, the polarization of the probe is along $x$ or $z$.  In TDTB measurements, the polarization of the probe is at an angle of 45$^\circ$ from the $x$ axis. 
}
\end{figure}

\subsection{Optical and Thermal Properties}

First, we discuss measurements of refractive index, electrical conductivities, heat capacity, and thermal conductivities of \ce{Fe2As}.
We use the refractive index to describe the optical properties of the material and to relate thermo-reflectance and thermo-birefringence.
We measure the heat capacity and decompose it into electron, phonon, and spin contributions;
this allows us to compare the magnetic heat capacity to temperature dependent TDTB.
From the heat capacity and thermal conductivity, we model the time-evolution of the temperature excursion created by the pump optical pulse.
Finally, the combination of the measured electrical conductivity and the Wiedemann Franz law allows us to separate the electronic and lattice contributions to the total thermal conductivity.

The equivalence of the $x$-axis and $y$-axis was demonstrated in prior work by neutron diffraction \cite{Fe2Asstructure,phasediagramFe2As} and torque magnetometry \cite{AchiwaTorque}.  First-principles density functional theory~(DFT) calculations give the ground-state lattice parameters as
$a=3.624\text{\,\AA}$ and $c=5.860\text{\,\AA}$,
 within 2\% of powder X-ray diffraction measurements at room temperature, 
$a=3.628\text{\,\AA}$ and $c=5.978\text{\,\AA}$.
The magnetic unit cell used in the calculation is twice as long in the $z$ direction.
The N\'eel vector in ground-state DFT calculations is oriented along the $x$ direction of the lattice. We confirmed the easy-plane magnetic structure by measuring the temperature dependent magnetic susceptibilities along the $x$ and $z$ crystallographic directions with a vibrating sample magnetometer. 

In the absence of magnetic order, the dielectric tensor of a tetragonal crystal is isotropic in the $x$-$y$ plane. However, if the N\'eel vectors have a preferred direction in the $x$-$y$ plane, the dielectric function is anisotropic on length scales smaller or comparable to the characteristic size of the magnetic domains. We expect that different magnetic domains are approximately randomly oriented along the $x$ and $y$ directions and that our laser beam size is large compared to the domain size. Therefore, the dielectric function we measure in the $x$-$y$ plane is isotropic.

An effective isotropic refractive index of \ce{Fe2As} was measured by ellipsometry of the (001) and (010) faces of the crystal. Immediately prior to the ellipsometry measurements, which take place under ambient conditions, we removed surface oxides and contaminants using argon ion beam milling. The effective isotropic refraction index is $ {n}=2.9+i3.3$ at a wavelength of $\lambda = 780$ nm. The optical reflectance calculated from this index of refraction is 0.56. The measured optical reflectance for both the (001) and (010) surfaces of the crystal at normal incidence and $\lambda = 780$ nm is 0.50. The optical absorption depth, $\lambda/(4\pi k)$, is 19 nm.  The refractive index, computed using DFT, for light polarized along the $x$, $y$ and $z$-axis of the crystal is $ {n}=\sqrt{ {\epsilon}}=4.295+i3.496$, $4.300+i3.501$ and $3.381+i4.039$ at 0~K, with a reflectance of 0.573, 0.574 and 0.619, respectively. 

%

The electrical resistivity of a polycrystalline sample of \ce{Fe2As} was reported previously as $\approx 220$~\si{\mu\Omega.cm}  at $T=300$~K \cite{zocco2012high}. The electrical resistivity has a shallow maximum near room temperature and decreases to $\approx 125$~\si{\mu\Omega.cm}  at $T=1$ K. We confirmed a value of electrical resistivity of our samples near room temperature of $240$~\si{\mu\Omega.cm} and the residual resistivity ratio at 300 K of 1.7. The stoichiometry of \ce{Fe2As} was evaluated using Rutherford backscattering spectrometry and Rietveld refinements to synchrotron X-ray and neutron diffraction data (see Supplemental Materials). These measurements converge on a slight Fe deficiency of 0.05 to 0.08 out of 2.
This value also agrees with the nominal Fe:As ratio used during synthesis (1.95:1). (Nominally 2.00:1 samples exhibit metallic Fe impurities.) The high concentration of Fe vacancies in \ce{Fe2As} likely causes the large residual resistivity.

\begin{figure}[hbt]
\centering
\begin{subfigure}{0.4\textwidth}
\centering
\includegraphics[width=\textwidth]{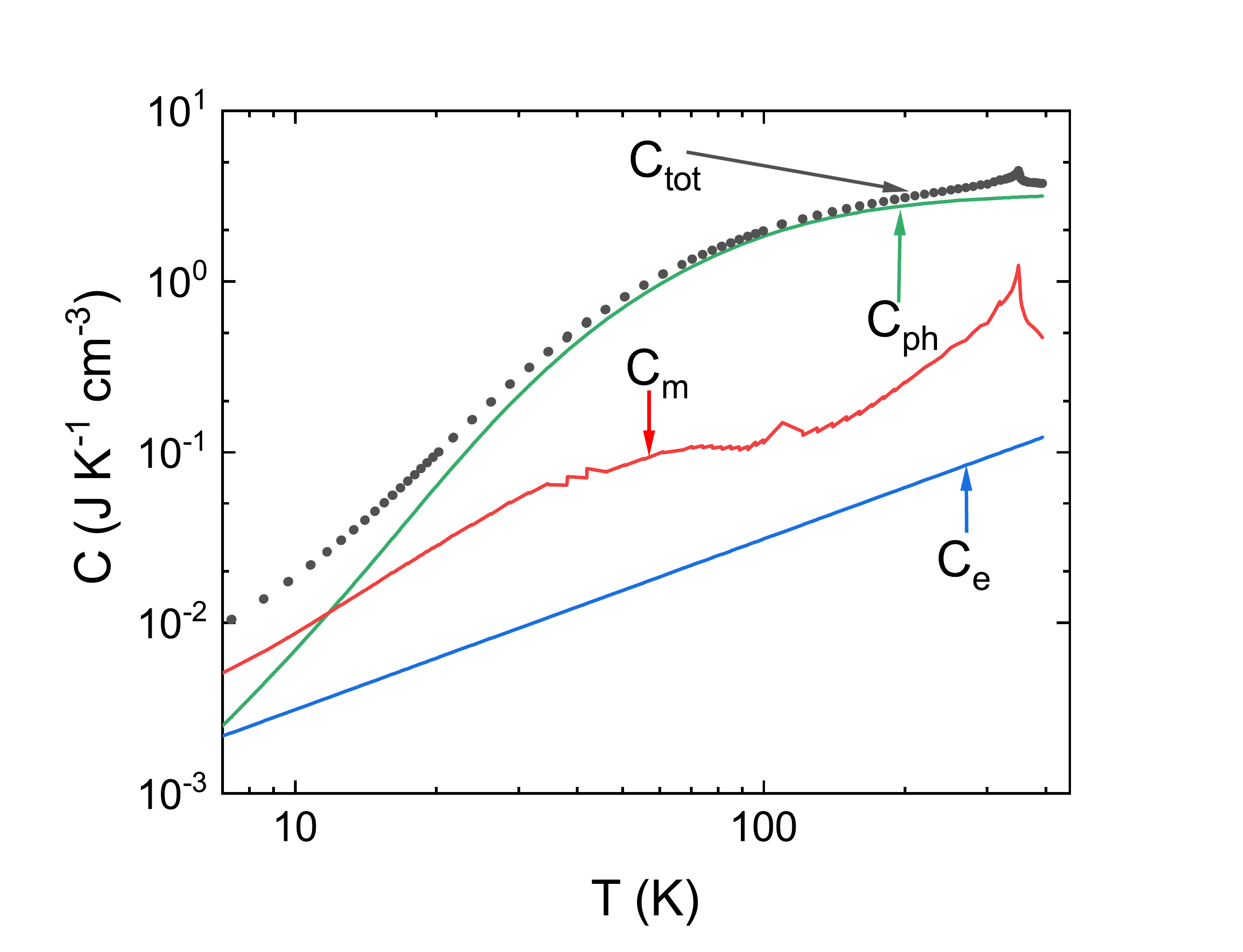}
\caption[]%
{{\small }}    
\label{fig:heat_capacity}
\end{subfigure}
\begin{subfigure}{0.4\textwidth}  
\centering 
\includegraphics[width=\textwidth]{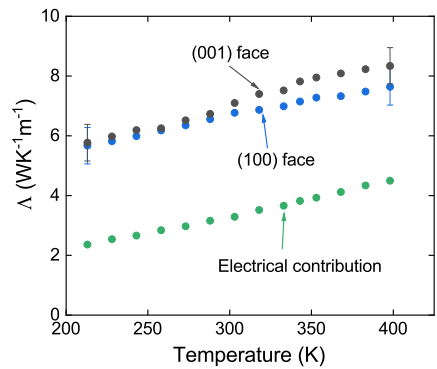}
\caption[]%
{{\small}}    
\label{fig:TC_200}
\end{subfigure}
\caption[]
{\small Heat capacity and thermal conductivity of \ce{Fe2As}. (a) The measured total heat capacity $C_{\rm tot}$ of \ce{Fe2As} and contributions to $C_{\rm tot}$ from excitations of electrons (e), phonons (ph), and magnons (m). The electronic and phonon contributions are calculated by density functional theory (DFT).  The magnon contribution is derived by subtracting the calculated phonon and electronic contributions from $C_{\rm tot}$ . (b) The  thermal conductivity in the direction normal to the (001) face (black circles) and (100) face (blue circles) shows a small anisotropy.  The electrical contribution to the thermal conductivity (green circles) is calculated from the Wiedemann-Franz law and measurements of the electrical conductivity. } 
\label{fig:TC_and_HC}
\end{figure}

\begin{figure}[htb]
\centering
\begin{subfigure}{0.4\textwidth}
		\centering
		\includegraphics[width=\textwidth]{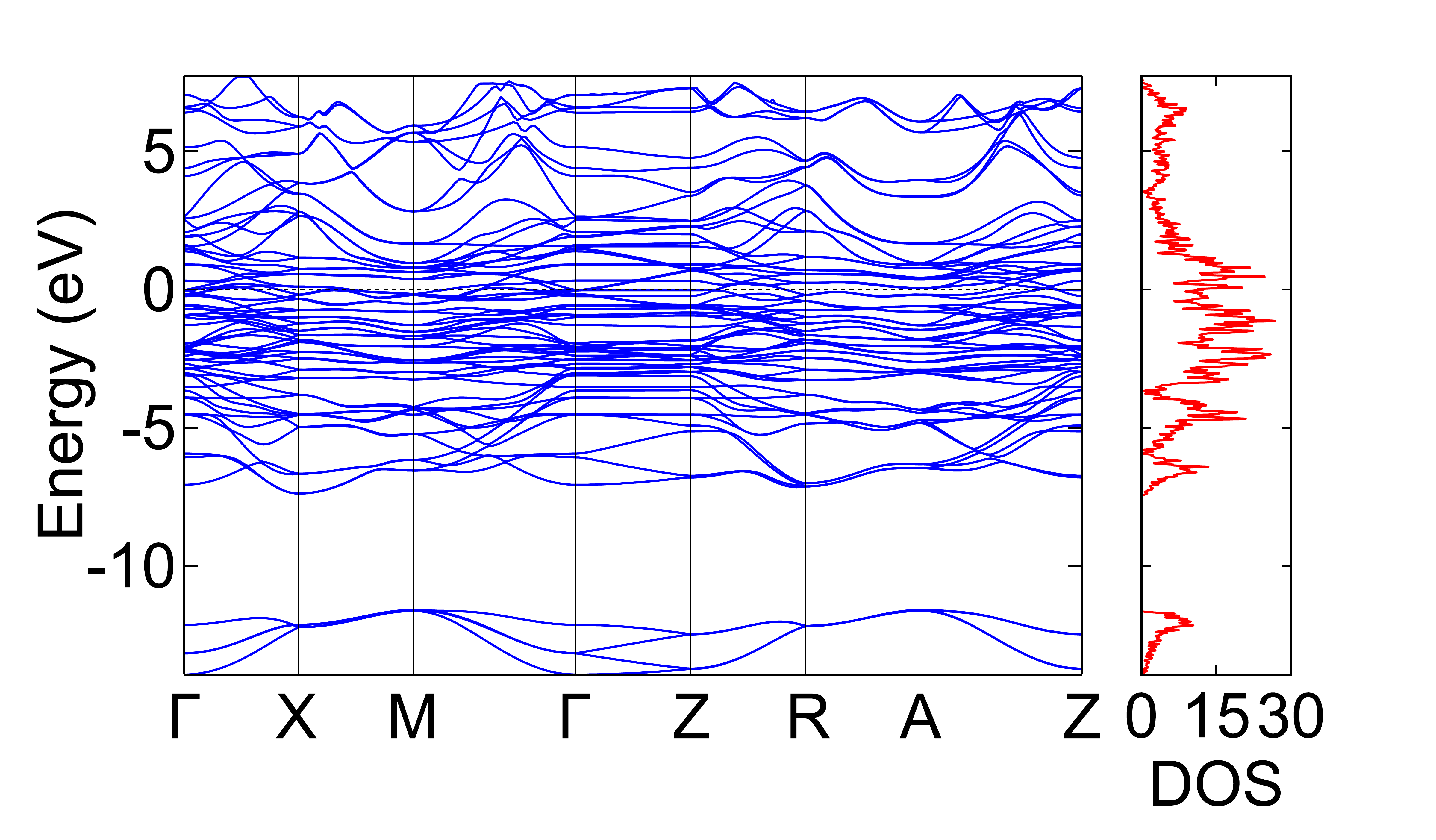}
		\caption[]
		{{\small }}   
		\label{fig:electron}
\end{subfigure}
\begin{subfigure}{0.4\textwidth}  
		\centering 
		\includegraphics[width=\textwidth]{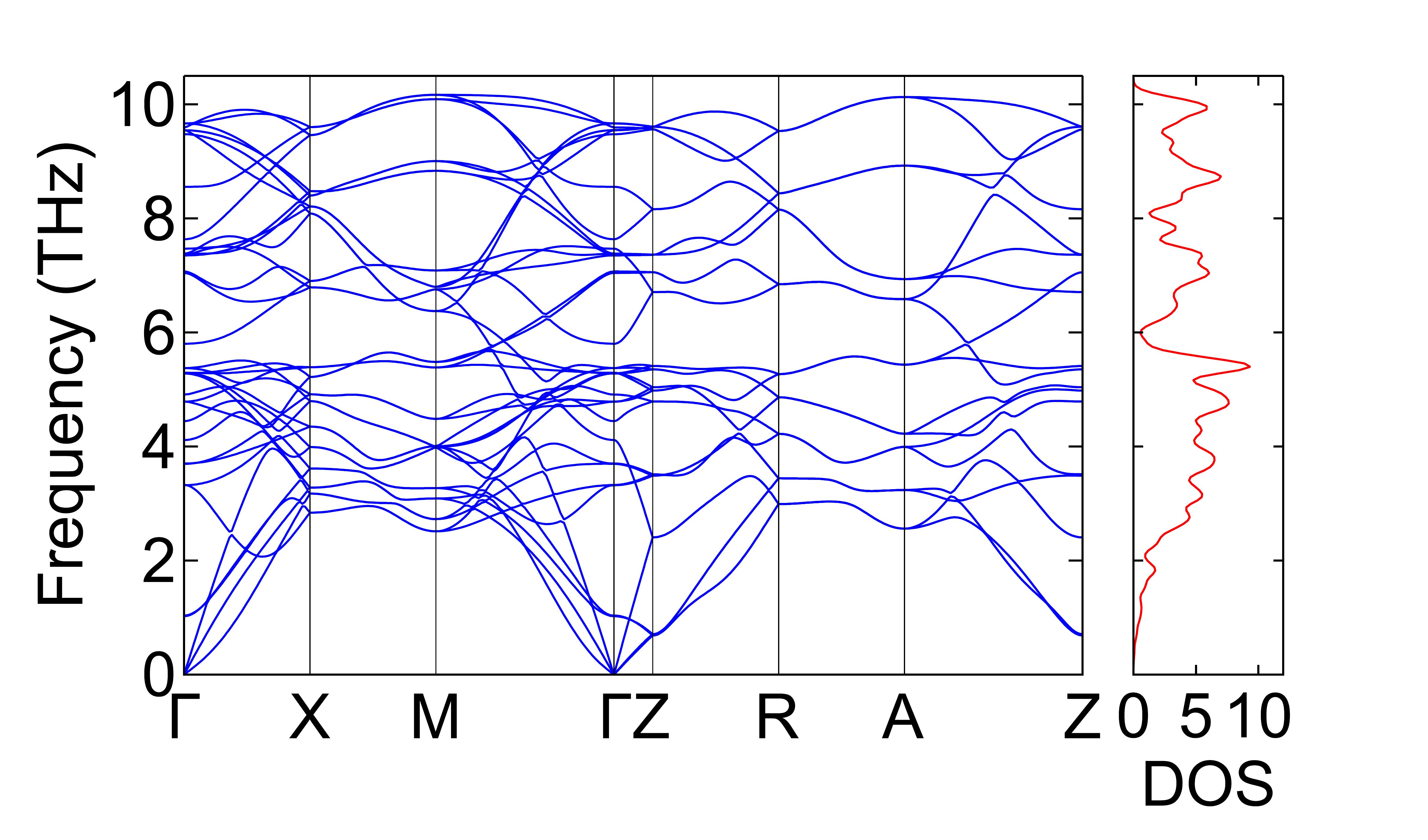}
		\caption[]
		{{\small }} 
		\label{fig:phonon}
\end{subfigure}
\caption[]
{
\label{fig:electron_and_phonon}(a) Calculated electronic band structure and electronic density of states (DOS) of \ce{Fe2As}. The electronic band structure include spin-orbit coupling effect through a non-collinear magnetism calculation. (b) Calculated phonon dispersion and phonon DOS of \ce{Fe2As}. For (a), the units of the electronic DOS are the number of states per magnetic unit cell per eV; for (b), the units of the phonon DOS are the number of states per magnetic unit cell per THz.
} 
\end{figure}

We measured the total heat capacity of a 35.5 mg sample of \ce{Fe2As} with a Quantum Design Physical Property Measurement System (PPMS), see Fig.\ \ref{fig:heat_capacity}.
The phonon heat capacity is calculated from the phonon density of states of the ground state crystal structure and magnetic ordering, as shown in Fig.\ \ref{fig:phonon}.
The Debye temperature calculated from the phonon DOS is 286~K, in good agreement with the measured value of 296~K reported by Zocco {\it et al.}\ \cite{zocco2012high}.

We computed the electronic density of states using Mermin DFT \cite{Mermin:1965} and finite electronic temperatures between 0 K and 400 K, see Fig.~\ref{fig:electron} for 300 K.
From this data we calculate the electronic heat capacity $C_{e}$ and electronic specific heat $\gamma$=7.41 \si{mJ.K^{-2}.mol^{-1}}. 

Finally, we assume that the magnetic heat capacity is given by the difference between the total heat capacity and the sum of the lattice  and electronic heat capacities, $C_m=C_{tot}-C_p-C_e$. Because lattice heat capacity dominates the total heat capacity except at very low temperatures,  small errors in the measurement of the total heat capacity or the calculation of the phonon heat capacity produce large uncertainties in the magnetic heat capacity.
We do not yet understand the origin of the small peak in the heat capacity data near 110~K.

We also measured the thermal conductivity of \ce{Fe2As} normal to the (001) and (010) faces of the crystal using conventional TDTR measurements and modeling as shown in Fig.\ \ref{fig:TC_200}. An 80 nm thick Al film was sputtered on the sample to serve as the optical transducer in the thermal conductivity measurement. We determined the thermal conductivity by fitting the data to a multilayer thermal model \cite{RSI_cahill}.
The electrical contribution to the thermal conductivity was estimated by using the combination of the Wiedemann-Franz law and the measured electrical resistivity.
The thermal conductivity shows a small anisotropy at $T>300$~K (see Fig.\ \ref{fig:TC_200}) and the contributions to the thermal conductivity from phonons and electronic excitations are comparable.
The phonon contribution, i.e.\ the difference between the measurement and the electronic contribution, is approximately 3.6~\si{W.K^{-1}.m^{-1}} and independent of temperature.

\subsection{Time-domain Thermo-birefringence and Time-domain Thermoreflectance}

In our experiments we probe changes in the optical function of \ce{Fe2As} induced by excitation of the sample by the pump beam.
We use time-domain thermoreflectance (TDTR) to measure changes in the diagonal components of the dielectric tensor by fixing the probe polarization along various crystallographic directions and measuring transient changes in the {\em intensity} of the reflected probe pulse. 
We use time-domain thermo-birefringence (TDTB) to measure changes in the difference between the diagonal components of the dielectric tensor through transient changes in the {\em polarization} of the reflected  probe pulse. For both TDTB and TDTR, the strongest signals we have observed are for pump and probe beams at normal incidence on the crystallographically anisotropic (010) surface of \ce{Fe2As}.

The instrument we used for the measurements of \ce{Fe2As} is the same as the instrument that we use for conventional TDTR measurements of thermal transport and time-resolved magneto-optical Kerr effect (TR-MOKE) of magnetization dynamics and thermally-driven spin generation and transport \cite{liu2014measurement, kimling2017thermal}. A Ti:sapphire laser oscillator with an 80 MHz repetition rate is separated into a pump beam and a probe beam. The pump beam is modulated at $\approx 11$ MHz, and the probe beam is modulated at 200 Hz. A translation stage is used to change the delay time $t$ between pump and probe optical pulses.

\begin{figure}[htb]
\centering
\begin{subfigure}{0.4\textwidth}
\centering
\includegraphics[width=\textwidth]{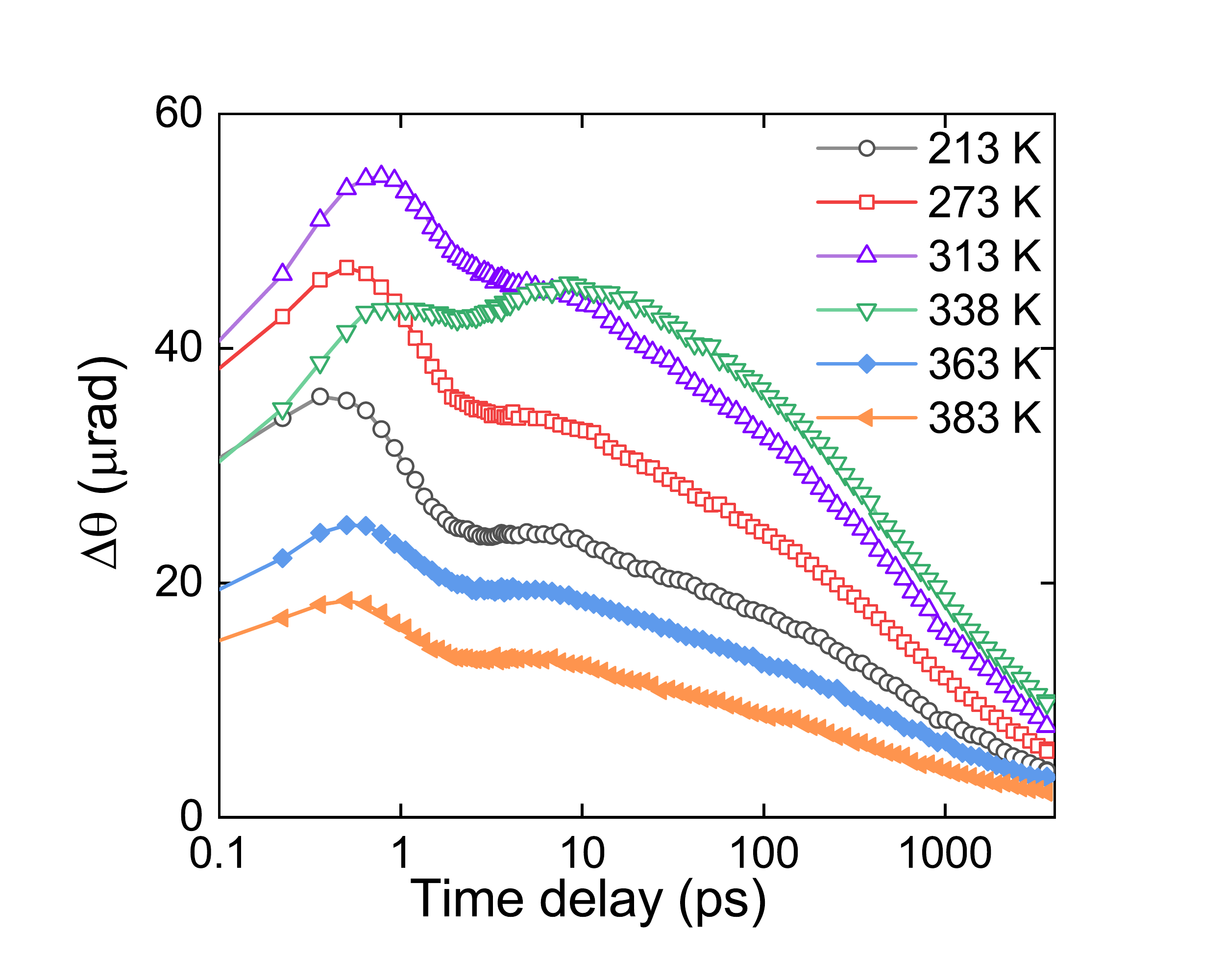}
 \caption[]
 {{\small }}   
\label{fig:TRQMOKE}
 \end{subfigure}
\begin{subfigure}{0.4\textwidth}  
\centering 
\includegraphics[width=\textwidth]{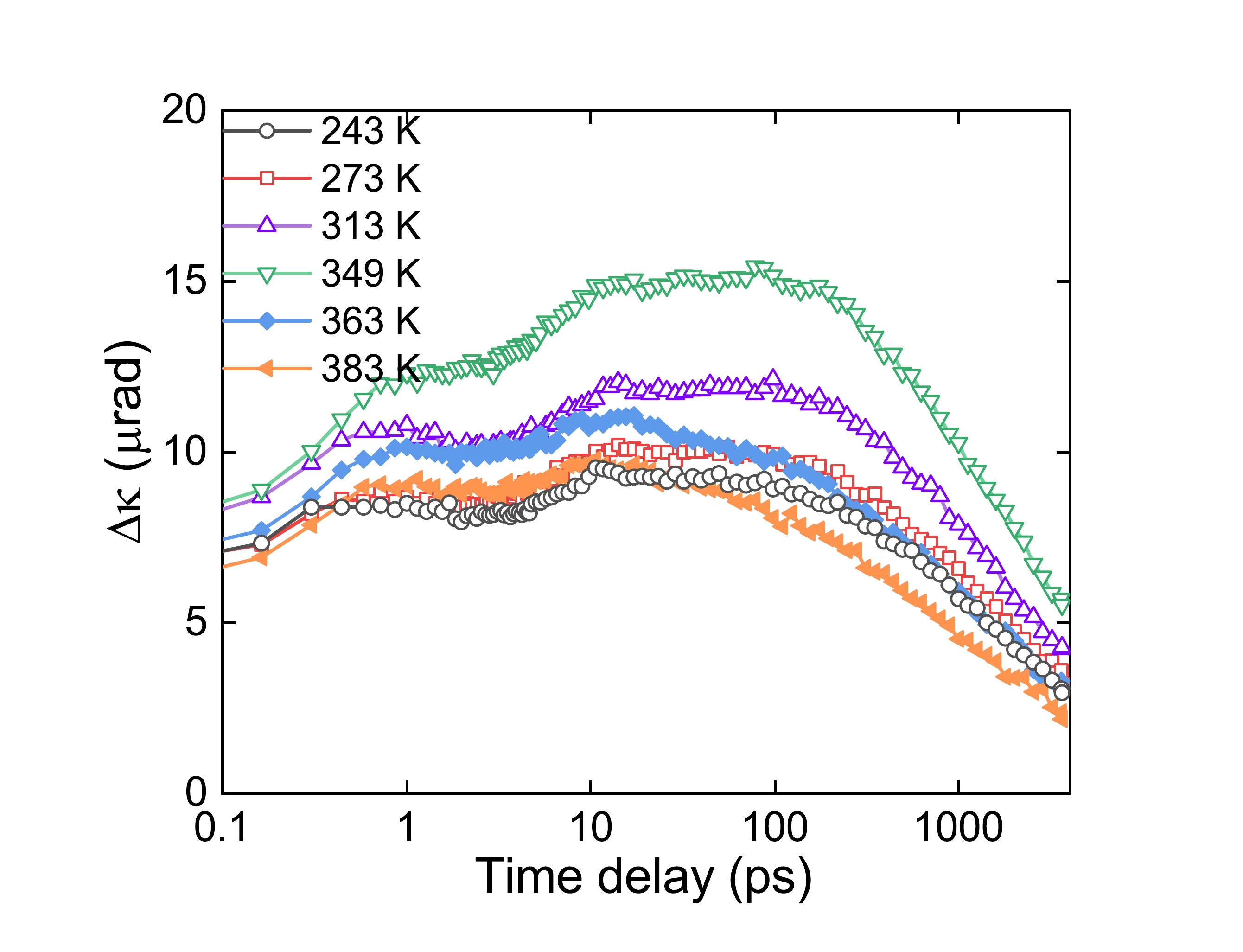}
 \caption[]
{{\small }} 
\label{fig:TR_KE}
\end{subfigure}
\caption[]
{\small (a) The real part of the time-domain thermo-birefringence (TDTB) signal measured on the (010) face of \ce{Fe2As}; and (b) the imaginary part of the TDTB signal. The temperature in the legend is the  temperature of the sample stage; the spatially averaged temperature of the area of the sample that is measured in the TDTB experiment is the sum of stage temperature and the steady-state heating of 13 K. When stage temperature is at 338 K, the temperature of measured region of the sample is close to $T_N=350$ K. Empty symbols denote data acquired at $T<T_N$; filled circles are data for $T>T_N$. We attribute the slower response at $T \approx T_N$ to the peak in the magnetic heat capacity at $T_N$. } 
\label{fig:KR_and_KE}
\end{figure}

\begin{figure}[htb]
	\centering
\begin{subfigure}{0.4\textwidth}   
\centering 
\includegraphics[width=\textwidth]{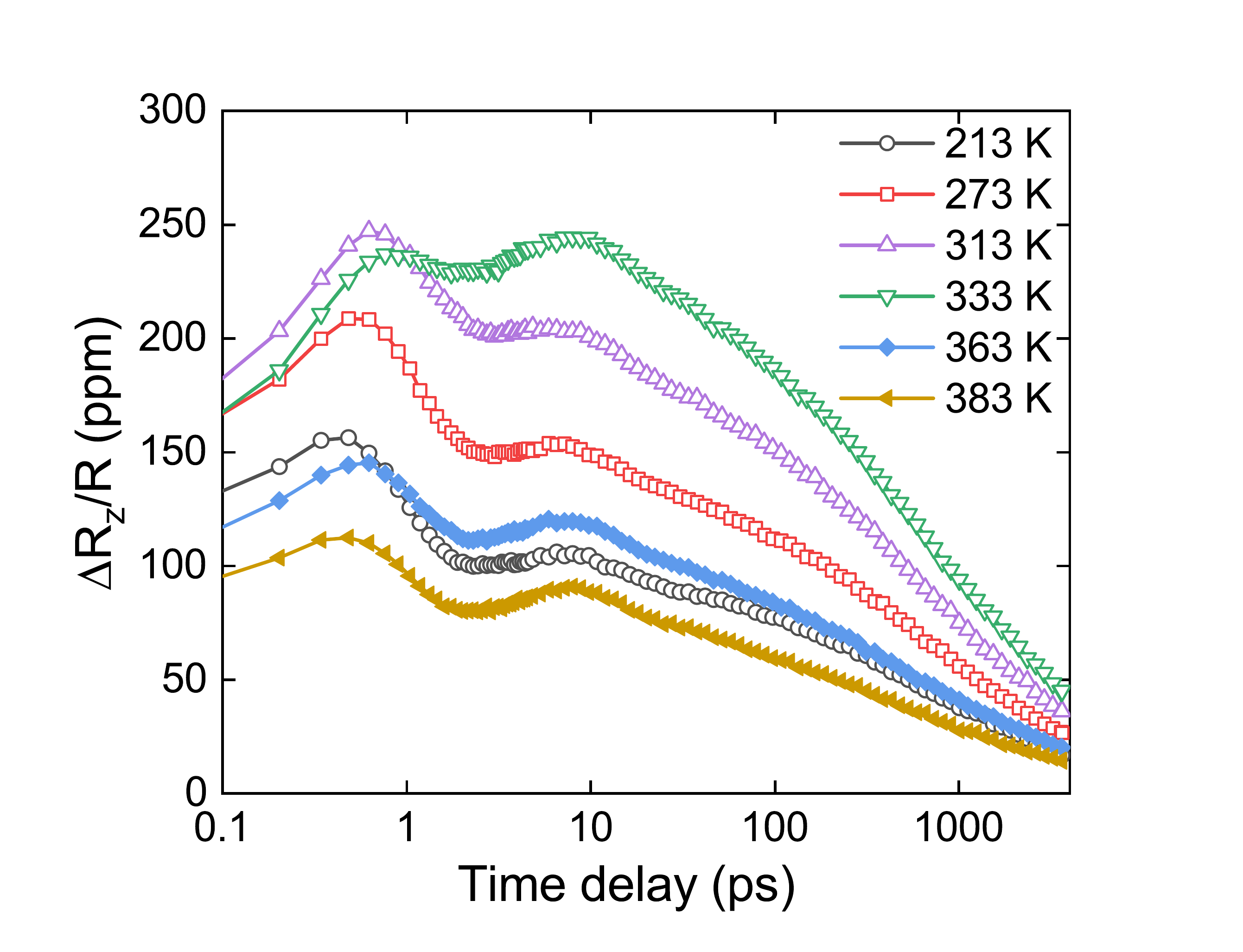}
\caption[]%
{{\small }}    
\label{fig:TDTR_z}
\end{subfigure}
\\
\begin{subfigure}{0.4\textwidth}   
\centering 
\includegraphics[width=\textwidth]{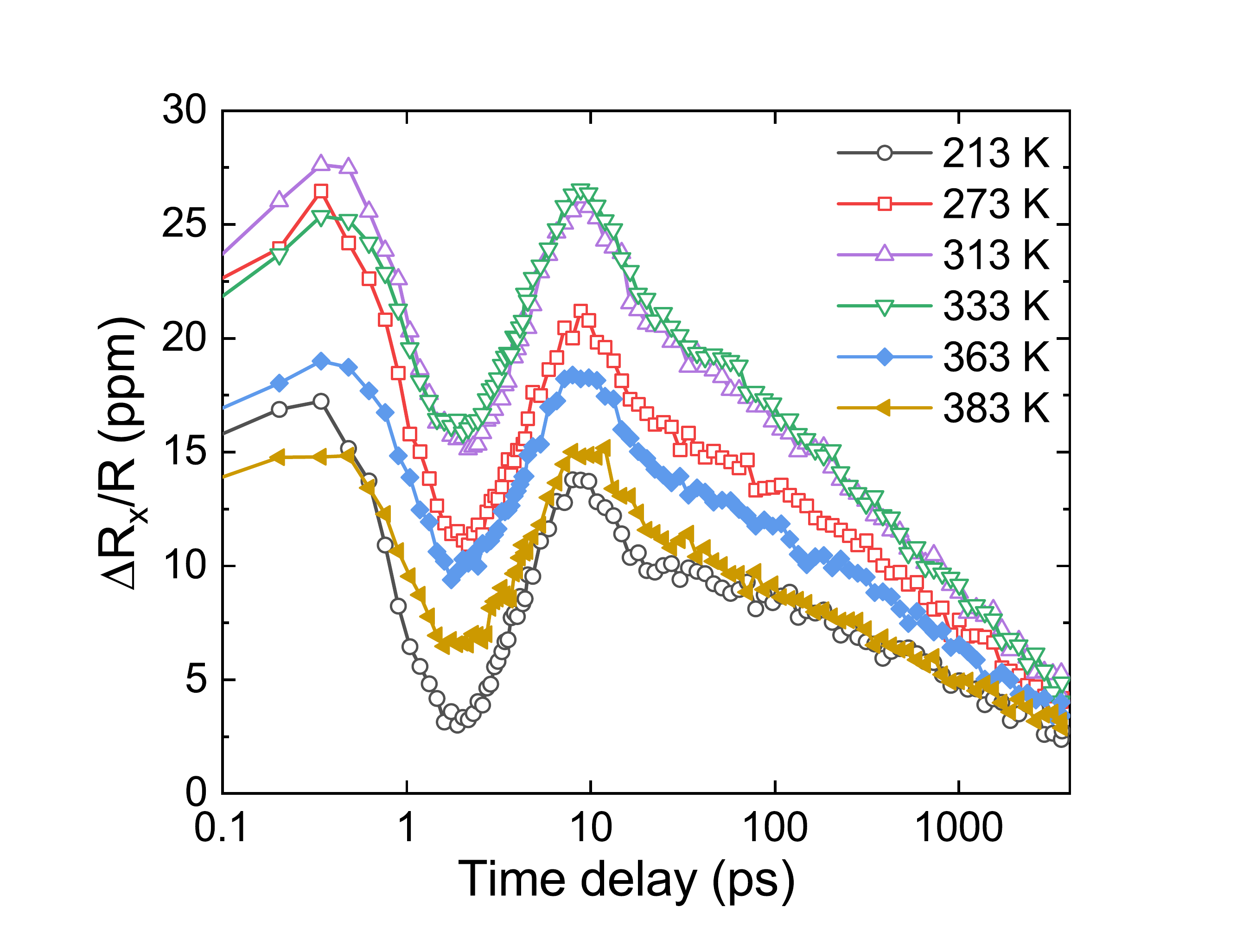}
\caption[]%
{{\small }}    
\label{fig:TDTR_x}
\end{subfigure}
\caption[]
{\small (a) Time domain thermoreflectance (TDTR) data for the (010) face of \ce{Fe2As} with (a) probe polarization aligned along the $z$-axis and (b) probe polarization aligned along the $x$-axis. The TDTR data for $\Delta R_z/\Delta T$ shown in panel (a) is approximately an order of magnitude larger than the TDTR data for $\Delta R_x/\Delta$ shown in panel (b). The temperature in the legend is the  temperature of the sample stage; the spatially averaged temperature of the area of the sample that is measured in the TDTR experiment is the sum of stage temperature and the steady-state heating of 13 K. Empty symbols denote data for temperatures $T<T_N$; filled symbols are for data acquired at $T>T_N$. } 
\label{fig:TDTR}
\end{figure}

The TDTB measurement geometry is shown in Fig.~\ref{fig:expr_diagram} and the  measurement results for the \ce{Fe2As} (010) face are shown in Fig.~\ref{fig:TRQMOKE}. In the discussion that follows, the symbol $\Delta$ indicates a transient quantity. To measure transient changes in the real part of the polarization rotation, $\mathrm{Re}[\Delta\Theta]=\Delta\theta$, we null the balanced detector with a half-wave plate before the Wollaston prism that splits the orthogonal polarizations into two paths that are focused onto the two photodiodes of the balanced detector; to measure transient changes in the imaginary part of the rotation, i.e., the ellipticity $\mathrm{Im}[\Delta\Theta]=\Delta\kappa$, we null the balanced detector with a quarter-wave plate.  The polarization of the probe-beam is in the $x$\,--\,$z$ plane and 45$^\circ$ from the $x$ axis.
Corresponding TDTR data for a bare \ce{Fe2As} (010) face, i.e.\ uncoated by a metal, is shown for the two orthogonal polarizations in Fig.~\ref{fig:TDTR_z}.

Optical reflectance is the ratio of the intensity of the reflected electrical field to the intensity of the incident electric field: $R_z=| {r}_{zz}|^2$  and $R_x=| {r}_{xx}|^2$ where $ {r}_{zz}$ and $ {r}_{xx}$ are the Fresnel reflection coefficients for light polarized along the $z$ and $x$ directions, respectively. $ {r}_{ii} = ( {n}_{ii}-1)/( {n}_{ii}+1)$ where $ {n}_{ii}$ are diagonal elements of the optical index of refraction tensor; $ {n}_{ii}^2 =  {\varepsilon}_{ii}$ where $ {\varepsilon}_{ii}$ are diagonal elements of the dielectric tensor. 

The birefringence of \ce{Fe2As} is relatively small. We therefore define average quantities  $\bar{n}=(n_{xx}+n_{zz})/2$; $\bar{\varepsilon}=\bar{n}^2$; and $\bar{r} = (\bar{n}-1)/(\bar{n}+1)$.
The complex rotation of the polarization of the reflected probe light is then
\begin{equation} \label{eq:rotation}
\begin{split}
\Theta& \approx \frac{\left( {r}_{xx}- {r}_{zz}\right)}{2 \bar{r}}
\approx \frac{( {n}_{zz}- {n}_{xx})}{(1- \bar{n}^2)} \approx \frac{( {\varepsilon}_{zz}- {\varepsilon}_{xx})}{2\sqrt{\bar{\varepsilon}}(1- {\bar\varepsilon})}.
\end{split}
\end{equation}
We use Eq.\ \ref{eq:rotation} to relate the polarization rotation angle to differences in the index of refraction or differences in the dielectric function. We evaluate Eq.~\ref{eq:rotation} using the measured refractive index $n=2.9+i3.3$. The real and imaginary parts of the TDTB signal can then be written as $\Delta\theta=0.005(\Delta\varepsilon'_{zz}-\Delta\varepsilon'_{xx})-0.003(\Delta\varepsilon''_{zz}-\Delta\varepsilon''_{xx})$ and $\Delta\kappa=0.003(\Delta\varepsilon'_{zz}-\Delta\varepsilon'_{xx})+0.005(\Delta\varepsilon''_{zz}-\Delta\varepsilon''_{xx})$, where $\varepsilon'_{ii}$ and $\varepsilon''_{ii}$ are real and imaginary parts of the relative dielectric tensor. 

Because the reflectance $R_i$ is a function of complex dielectric function $ {\varepsilon}_{ii}$,  the transient reflectance, or TDTR signal, can be written as $\Delta R=\frac{\partial R}{\partial\varepsilon'}\Delta \varepsilon'+\frac{\partial R}{\partial \varepsilon''}\Delta\varepsilon''$.
After taking partial derivative of reflectance and inserting the static dielectric function calculated from the measured refractive index, the transient reflectance can be written as a linear combination of transient dielectric functions, $\Delta R_i=-0.01\Delta\varepsilon'_{ii}+0.007\Delta\varepsilon''_{ii}$.


We note that the difference in the TDTR measurements along $x$ and $z$  closely resembles the real part of the TDTB signal, $\Delta \theta$.  This is because the linear coefficients of the transient changes in the components of the dielectric tensor that contribute to $\Delta R_{xx} - \Delta R_{zz}$ are approximately twice the linear coefficients of the transient changes in the components of the dielectric tensor that contribute to $\Delta\theta$. In other words, $\Delta\theta\approx-(\Delta R_z-\Delta R_x)/2$. 
Alternatively, if we write the complex TDTB signal as an amplitude and phase in the form $\Delta\Theta=|z|e^{i\delta}$, the real part of the TDTB signal is $\Delta\theta=|z|\cos{\delta}$, while in TDTR measurement, $\Delta R_z-\Delta R_x=2|z|$.

The per pulse heating, i.e., the temperature excursion produced by a single optical pulse of the pump beam, is $\Delta T(t) \approx 3$~K. Due to the small temperature excursion, the change in  the sublattice magnetization  $\Delta M$ is small compared to the sublattice magnetization $M$, except for $T$ very close to $T_N$. This justifies a description of the experiment in terms of linear response, except for $T$ very close to $T_N$.

Since the (010) face of a tetragonal crystal is fundamentally anisotropic, we cannot directly interpret the signals plotted in Fig.~\ref{fig:KR_and_KE} as the result of changes in magnetization with temperature. 
There are, however, two aspects of the data that suggest a prominent role of magnetism and magneto-optic effects.
First, the real part of the TDTB signal (see Fig.\ \ref{fig:TRQMOKE}), and the TDTR signal measured with the probe polarization along the $z$ axis  (see Fig.\ \ref{fig:TDTR_z}),  show a significantly slower response when the sample temperature is close to the N\'eel temperature, $T_N$
The transfer of thermal energy in a magnetic material is often described by a three temperature model, in which energy is transferred between electrons, phonons, and magnons on ultrafast time-scales \cite{Bigot_Ni,Kimling2014}. Since the magnon contribution to heat capacity reaches a maximum at $T_N$, the temperature rise of the magnon system in response to heating of the electronic system by the pump optical pulse is expected to be slower at temperatures near $T_N$. We attribute the slower optical response at $T\approx T_N$ to this effect and conclude that the real part of the TDTB signal, and the TDTR signal measured with polarization along the $z$-axis, are dominated by changes in the magnon temperature.  The slowing down of the demagnetization of antiferromagnetic \ce{Fe2As} at $T \approx T_N$ is reminiscent of the slowing down of the demagnetization of ferromagnetic FePt:Cu at $T\approx T_C$ where $T_C$ is the Curie temperature \cite{Kimling2014}.

\begin{figure}[hbt]
\centering
\includegraphics[width=0.4\textwidth]{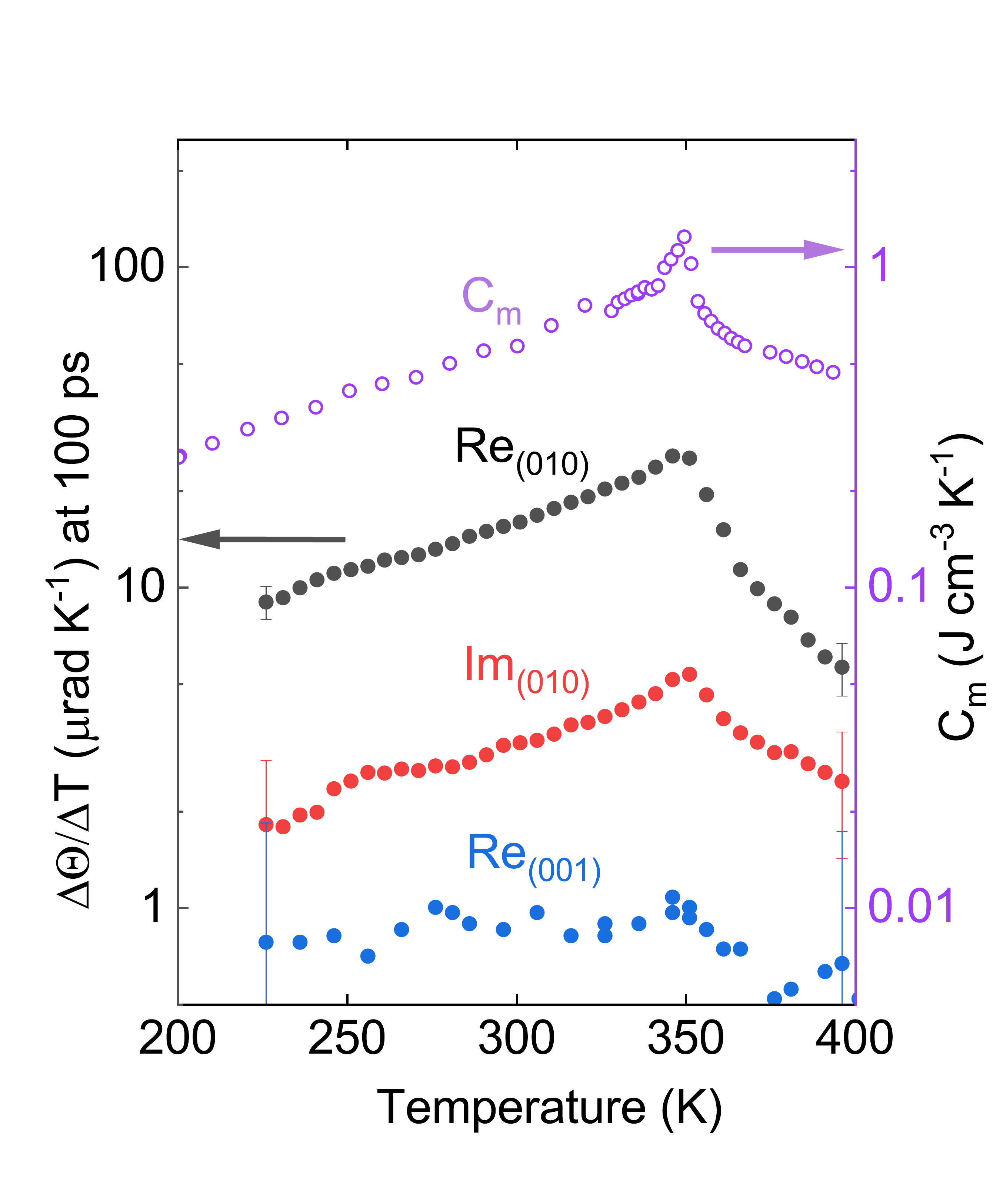}
\caption{\label{fig:MOKE_vs_T_2} Comparison between  $\Delta\Theta/\Delta T$ and magnetic specific heat as a function of sample temperature. The values for $\Delta\Theta$ are for 100 ps delay time. For each sample temperature $T$, $\Delta T$ at 100 ps at  is calculated from a thermal model that uses the measured total heat capacity and thermal conductivity of \ce{Fe2As} as inputs to the model. The sample temperature $T$ includes the effects of steady-state heating of measurement area that is created by the absorbed laser power. The real and imaginary parts $\Delta\Theta/\Delta T$ measured for the (010) face have a similar temperature dependence as the magnetic specific heat $C_m$.}
\end{figure}

Second, the temperature dependence of the magnitude of the transient TDTB and TDTR signals at fixed time delays closely follows the magnetic heat capacity. In Fig.~\ref{fig:MOKE_vs_T_2}, we compare the complex thermo-birefringence signals $\Delta\Theta/\Delta T$ for the (010) plane of \ce{Fe2As} and the magnetic heat capacity derived from $C_m=C_{tot}-C_{ph}-C_e$.  We use TDTB data acquired at pump-probe delay times near 100 ps when the electrons, magnons, and phonons are in thermal equilibrium and the strain and temperature gradients within an optical absorption depth of the surface are small. As we discuss in more detail below, we expect that for a single mechanism, the magnetic contribution to the dielectric function of an antiferromagnetic material will scale with the magnetic energy and, therefore, transient changes in the dielectric function that are produced by a small temperature excursion will scale with the magnetic heat capacity $C_m$. At $T\approx T_N$, we expect that $\Delta\Theta/\Delta T$ will be more smoothly varying with $T$ than $C_m$ because of the inhomogeneous temperature distribution across the lateral extent of the pump and probe beams in the TDTB experiment.

In Fig.~\ref{fig:MOKE_vs_T_2} we also include the temperature dependence of $\Delta \theta$ measured on the crystallographically isotropic (001) plane. We consistently observe a small signal that is approximately independent of position.  We believe there are two mechanisms that contribute to this null result.  For the (001) plane of \ce{Fe2As}, the two degenerate domain orientations should produce  a cancellation of any TDTB signal when measured on a length scale large compared to the characteristic domain size. We have not yet determined the domain structure of our \ce{Fe2As} crystals but evidence from related materials \cite{grzybowski2017imaging,sapozhnik2018direct} suggest that the domain size is typically in the sub-micron range while the $1/e^2$ radius of the pump and probe laser beams is $\approx 5.5~\mu$m. Furthermore, the lack of a significant TDTR signal for light polarized along the $\left<100\right>$ of the (010) face suggests that magnetic contributions to the $\varepsilon_{xx}$ and $\varepsilon_{yy}$ components of the dielectric tensor are small. We tentatively attribute the small Kerr rotation signal that we observe on the (001) face to a small, uncontrolled miscut of the sample, i.e., a small misorientation between the surface normal and the $c$-axis of the crystal.

Since both the real $\Delta\theta$ and imaginary  $\Delta\kappa$ parts of the TDTB signals measured on the (010) face have a temperature dependence that resembles the magnetic heat capacity, we conclude that both $\Delta\theta$ and $\Delta\kappa$ have significant magnetic contributions. However, $\Delta\theta$ and $\Delta\kappa$ do not have the same dynamics, see Fig.~\ref{fig:KR_and_KE}. In the $\Delta \theta$ data set, with the exception of data collected at $T\approx T_N$, the signal reaches a peak response at short delay times on the order of 1 ps. We interpret this signal as arising from the same type of out-of-equilibrium ultrafast demagnetization that is typically observed for ferromagnetic materials using pump-probe measurements of the polar or transverse magneto-optic Kerr effects. However, we cannot yet reliably distinguish between magnetic, electronic, lattice temperatures, and lattice strain contributions to $\Delta \theta$ or $\Delta\kappa$.

In the $\Delta\kappa$ data set, the signal reaches a peak response on a time-scale on the order of 10 ps. We interpret this time scale as characteristic of the time needed to fully relax the thermoelastic stress within the near surface region of the crystal that determines the reflection coefficients of the probe beam.  This interpretation is supported by the character of the TDTR signal measured on the (001) face, see Fig.~\ref{fig:MOKE_vs_T_2}, that also includes a large variation in the signal at $t<20$~ps.

In most studies of the optical properties of materials, the thermal expansion of the material contributes to the temperature dependence of the dielectric tensor. Our experiments take place in a different regime. Thermal stress is generated when the pump optical pulse is partially absorbed by the near-surface region of the sample.  Thermal strain in the in-plane direction is strongly suppressed in a modulated pump-probe experiments because the thermal penetration depth, i.e., the depth of the heated region at the frequency of the pump modulation, is small compared to the lateral extent of laser spot. 

On the other hand, strain in the out-of-plane direction can contribute to the TDTB and TDTR signals. The probe beam  is sensitive to the dielectric tensor of the near-surface layer of the crystal that lies within an optical absorption depth of the surface. On this length scale, strain normal to the surface evolves on a time scale given by the optical absorption depth divided by the longitudinal speed of sound. The longitudinal speed of sound from our DFT calculations is $\approx 5$ nm/ps. Therefore, the  characteristic time-scale is $\approx 4$ ps. At $t\ll 4$ ps,  {\em strain} normal to the surface is negligible; at $t \gg 4$ ps, {\em stress} normal to the surface is negligible.  On long time scales, the decay of the strain normal to the surfaces will follow the decay of the surface temperature as heat diffuses into the bulk of the sample.

\section{Discussion}

TDTB and TDTR signals are linearly related to transient changes in the dielectric function, see Eq.~\ref{eq:rotation}. The dielectric function tensor can be written as \cite{pisarev1971magnetic} 
\begin{equation}
\varepsilon_{ij}=\varepsilon^0_{ij}+K_{ijk}M_k+G_{ijkl}M_kM_l+...
\end{equation}
where the first term is the non-magnetic contribution to $\varepsilon_{ij}$, and $K_{ijk}$ and $G_{ijkl}$ are the first- and second-order magneto-optic coefficients. Because the net magnetization is zero in a collinear AF, the linear term can be neglected here.  In the following discussion, we assume an equal population of magnetic domains with N\'eel vectors in the $x$ and $y$ directions. The magnetic contributions to the dielectric tensor are 

\begin{equation} \label{eq:G-tensor}
\begin{split}
&\varepsilon_{11}= G_{11}\sum_{i,j}\langle  S^i_x S^j_x \rangle+G_{12}\sum_{i,j}\langle  S^i_y S^j_y \rangle+\varepsilon^0_x\\
&\varepsilon_{22}=G_{11}\sum_{i,j}\langle  S^i_y S^j_y \rangle+G_{12}\sum_{i,j}\langle  S^i_x S^j_x \rangle+\varepsilon^0_y\\
&\varepsilon_{33}=G_{31}\left(\sum_{i,j}\langle  S^i_x S^j_x \rangle+\sum_{i,j}\langle  S^i_y S^j_y \rangle\right)+\varepsilon^0_z
\end{split}
\end{equation}
We have adopted the Voigt notation with  $G_{11}$, $G_{12}$ and $G_{31}$ replacing $G_{1111}$, $G_{1122}$ and $G_{3311}$, respectively. For tetragonal symmetry, $G_{22}=G_{11}$, $G_{12}=G_{21}$ and $G_{32}=G_{31}$. In Eq.~\ref{eq:G-tensor}, the labels $i$ and $j$ refer to  sites on the magnetic sublattice within the magnetic unit cell, the notation $S_x^i$ denotes the $x$ component of the sublattice magnetization of the $i$th lattice site, and the brackets $\langle \rangle$ denote an average. We assume that there is no correlation in the sublattice magnetization along the $z$ axis and therefore terms that involve $S_z$ average to zero. 

The microscopic mechanisms that contribute to the quadratic magneto-optic coefficients $G_{ijkl}$ can be categorized in terms of the exchange interaction, the Voigt effect, and magnetostriction \cite{ferre1984linear}. The Voigt effect only affects the value of $G_{11}$; the exchange interaction and magnetostriction contribute to all three coefficients $G_{11}$, $G_{12}$ and $G_{31}$. The magnetostriction contributions to $G_{ijkl}$ can be further divided into changes in the lattice parameters and changes of the atomic positions within a unit cell \cite{jauch1991structural}.

In the TDTB measurement of (001) face, if the dimension of magnetic domains is larger than the laser spot, only the Voigt effect term is left because contributions from the exchange interaction should be isotropic in the $x$-$y$ plane, and contributions from magnetostriction are  small because the in-plane strain in the transient measurement is negligible. In this case, our TDTB signal for the (001) face can be expressed as $\Delta\theta_{(001)}=(G_{11}'^{\parallel}-G_{11}'^{\perp})\sum_{i,j}\Delta\langle S_x^i S_x^j \rangle$, with assumption that N\'eel vector is along $x$ direction and the prime notation distinguishes the value of $G_{11}'$ for plane stress condition of the transient measurement from the value of $G_{11}$ for mechanical equilibrium. If the domain size  is much smaller than the beam spot size, the TDTB signal cannot be observed because  contributions to the TDTB signal from  the two orthogonal domains will cancel each other. 

In the easy-plane of tetragonal AFs (or cubic AFs), the detection of quadratic magnetization can sometimes be used to image magnetic domains \cite{NiOimaging}, because the changes of Voigt terms and lattice parameter are  coupled to magnetic anisotropy. In a previous study with tetragonal CuMnAs \cite{CuMnAs}, the large quadratic MOKE signal on the $x$-$y$ plane demonstrates the usefulness of the Voigt effect for studying the magnetic domain structure of  tetragonal CuMnAs. We find, however, that the Voigt effect contribution to the magneto-optic coefficients is small for \ce{Fe2As} and we have not yet been able to use TDTB signals to study the magnetic domain structure.

On the (010) face, approximately one-half of the domains have  N\'eel vectors in the in-plane $x$ direction and approximately one-half of the domains have N\'eel vectors in the out-of-plane $y$ direction. (The $z$ direction of the crystal lies in the plane of the (010) face of the crystal.) Magnetic contributions to the TDTB signals are only sensitive to the in-plane magnetic domains, $\Delta\theta_{(010)}=(G_{11}'-G_{31}')\sum_{i,j}\Delta\langle S_x^i S_x^j \rangle$.   If the Voigt effect were dominant, then we would expect $G_{11}'\gg G_{31}'$ \cite{LeGall,tesavrova3Dtrajectory}. However, the TDTR data, see Fig~\ref{fig:TDTR}, leads us to conclude that $\Delta\varepsilon_{zz} \gg \Delta\varepsilon_{xx}$ and therefore $G_{11}'\ll G_{31}'$.  The Voigt effect does not affect the value of $G_{31}'$ and thus the microscopic origins of TDTB signals of \ce{Fe2As} are dominated by  exchange interaction or atomic displacements within the unit cell.

In previous studies of tetragonal transition metal fluorides \cite{ferre1984linear,jahn1971change}, the derivative of the magnetic birefringence with respect to temperature $d(\Delta n_m)/dT$ has been shown to have the same temperature dependence as the magnetic specific heat. This behavior is expected because both the magnetic contributions to the dielectric function and the magnetic energy are proportional to terms of the form $\langle  S^i S^j \rangle$.  Contributions to the magneto-optic coefficients from exchange interactions and magnetostriction could, however, have different constants of proportionality.  Furthermore, in our TDTB measurements of the (010) face of the \ce{Fe2As} crystal, there could be significant non-magnetic contributions to the TDTB signals.  The fact that our thermo-birefringence signals closely resemble the temperature dependence of the magnetic heat capacity supports our conclusion that the thermo-birefringence signals are dominated by a magnetic contribution with a single underlying mechanism.

\begin{table*}[t]
\caption{Comparison of the quadratic magneto-optical coefficient $G_{31}$ of antiferromagnetic \ce{Fe2As} determined in our work with selected previous studies of ferromagnetic (Fe, Co, Ni, \ce{Y3Fe5O12}) and anti-ferromagnetic (MnF$_2$, CoF$_2$)  materials.  Nickel and \ce{Fe2As} have relatively large quadratic magneto-optic coefficients.
}
\label{Tab:MO_coefficient}
\centering
\scalebox{0.95}{
\begin{ruledtabular}
\begin{tabular}{@{}lllll@{}}
Materials & Wavelength & Birefringence data& Magnetization & Quadratic magneto-optical coefficient  \\ 
& (nm) & &  &($10^{-14}$\si{A^{-2}.m^2}) \\ \hline

Fe \cite{Felarge}& 670 &$\varepsilon_{11}-\varepsilon_{12}=-(5.0+i3.5)\times 10^{-2}$&  $1.8\times 10^6$\si{A.m^{-1}} & $G_{11}-G_{12}=-1.5-i 1.1 $ \\

Fe \cite{FeCuNiQMOKE}& 670 &$\varepsilon_{11}-\varepsilon_{12}=-0.15+i0.07$&  $1.8\times 10^6$\si{A.m^{-1}} & $G_{11}-G_{12}=-4.6+i2.1$ \\

Co \cite{FeCuNiQMOKE}& 670 &$\varepsilon_{11}-\varepsilon_{12}=0.10-i0.13$&  $1.4\times 10^6$\si{A.m^{-1}} & $G_{11}-G_{12}=5.1-i6.6$ \\

Ni \cite{FeCuNiQMOKE}& 670 &$\varepsilon_{11}-\varepsilon_{12}=-0.75+i0.20$&  $5.0\times 10^5$\si{A.m^{-1}} & $G_{11}-G_{12}=(-300+i80)$ \\

\ce{Y3Fe5O12} \cite{pisarev1971magnetic} & 1150 &$|n_\perp-n_\parallel|=3.9\times 10^{-6}$& $1.4\times 10^5$\si{A.m^{-1}} & $G_{11}-G_{12}=1.6$ \\

 \ce{MnF2}\cite{jahn1973linear}& 632.8 & $d\Delta n_m/dT=5.0\times 10^{-5}$~\si{K^{-1}}  &$d(M^2)/dT$ & $G_{33}-G_{13}=-1.2$ \\
 &  &  &$=-1.3\times 10^{10}$~\si{A^{2}.m^{-2}.K^{-1}} & \\ 
 
 \ce{CoF2}\cite{jahn1973linear}& 632.8 &$d\Delta n_m/dT=2.5\times 10^{-5}$~\si{K^{-1}}  &$d(M^2)/dT$ & $G_{33}-G_{13}=-0.63$ \\
&  &  &$=-1.2\times 10^{10}$~\si{A^{2}.m^{-2}.K^{-1}} &  \\
 
 \ce{Fe2As}& 783 & $d(\varepsilon_{\perp}-\varepsilon_{\parallel})/dT$ & $d(M^2)/dT$ & $G_{31}=(170-i23)$\\
 && $=(-1.5+i0.21)\times 10^{-3}$~\si{K^{-1}} &$=-8.8\times 10^{8}$~\si{A^{2}.m^{-2}.K^{-1}}&\\
\end{tabular}
\end{ruledtabular}}
\end{table*}

If we assume that only the sublattice magnetization contribution to the diagonal component of dielectric function is temperature-dependent, the magneto-optical coefficient $G_{31}$ can be estimated by using the value of $\Delta {\varepsilon}_{zz}$ and $\Delta(M_a^2)$. We estimate $\Delta(M_a^2)$ from our magnetic heat capacity data as described in the supplementary document. 
Because the sublattice magnetization is always real, the magneto-optic coefficient $G_{31}$ is complex since $\Delta {\varepsilon}_{zz}$ is a complex number. The magneto-optical coefficient \begin{equation}G_{31}=\frac{\Delta {\varepsilon}_{zz}/\Delta T}{\Delta(M_a^2)/\Delta T}=\frac{(\Delta\varepsilon_{zz}'+i \Delta\varepsilon_{zz}'')}{\Delta(M^2_a)}.  \end{equation}
Inserting the value of  the transient dielectric function and the temperature excursion of 1.5 K at a delay time of 100 ps and ambient temperature, 293 K, we find $G_{31}=(1.7-i0.23)\cdot10^{-12}$\si{A^{-2}.m^2}.

Finally, we compare the magnitude of our result for $G_{31}$ of \ce{Fe2As} with the quadratic magneto-optic coefficients of several more commonly studied  magnetic materials, see Table~\ref{Tab:MO_coefficient}.  In studies of ferromagnetic materials (Fe, Co, Ni, and \ce{Y3Fe5O12}), the magnetization vector can be manipulated with an external field and therefore the components of the quadratic magneto-optic tensor $G_{ij}$ can be calculated using Eq.~\ref{eq:G-tensor} and $M^2=M^2_s$, where $M_s$ is the saturation magnetization. In studies of antiferromagnetic materials (\ce{MnF2}, \ce{CoF2}, and \ce{Fe2As}), typically, the N\'eel vector cannot be controlled with an external field and the values of $G_{ij}$ are more difficult to determine. The measurements of antiferromagnetic \ce{MnF2} and \ce{CoF2} reported in Ref.~\cite{jahn1973linear} are collected from a crystallographic anisotropic plane; therefore, the magnetic birefringence data that we use in this analysis are the temperature derivatives of the birefringence data with the additional assumption that the magnetic birefringence has a stronger temperature dependence than the crystalline birefringence. We used $M^2=M_a^2$ to calculate the $G_{ij}$ tensor for antiferromagnets, where $M_a$ is sublattice magnetization.
Typically,  $G_{11}$ and $G_{13}$ or $G_{12}$ cannot be determined separately based on birefringence data alone. Compared to the other materials listed in Table~\ref{Tab:MO_coefficient}, \ce{Fe2As} and Ni have relatively large quadratic magneto-optic coefficients. 

\section{Conclusion}
In collinear antiferromagnetic materials, the contribution to the diagonal components of the dielectric tensor that are quadratic in sublattice magnetization can be probed with transient birefringence or reflectance measurements. In our measurement of time-domain thermo-birefringence (TDTB) and thermo-reflectance (TDTR) of \ce{Fe2As},  we observe that the dominant response of the dielectric tensor is in the direction perpendicular to the N\'eel vector, counter to the conventional expectation based on the Voigt effect, which only contributes to the dielectric tensor in the direction parallel to the N\'eel vector.  The temperature dependent TDTB signals closely follow the magnetic heat capacity, as expected if the exchange interaction is the dominant magnetic contribution to the dielectric tensor. In comparison to other magnetic materials, \ce{Fe2As} has relatively large quadratic magneto-optical coefficient $G_{31}$ at 783 nm.
%
%
%
%

\section*{Acknowledgments}

This work was undertaken as part of the Illinois Materials Research Science and Engineering Center, supported by the National Science Foundation MRSEC program under NSF award number DMR-1720633. 
This work made use of the Illinois Campus Cluster, a computing resource that is operated by the Illinois Campus Cluster Program (ICCP) in conjunction with the National Center for Supercomputing Applications (NCSA) and which is supported by funds from the University of Illinois at Urbana-Champaign.
This research is part of the Blue Waters sustained-petascale computing project, which is supported by the National Science Foundation (awards OCI-0725070 and ACI-1238993) and the state of Illinois. Blue Waters is a joint effort of the University of Illinois at Urbana-Champaign and its National Center for Supercomputing Applications.
This research used resources of the Spallation Neutron Source, a DOE Office of Science User Facility operated by  Oak Ridge National Laboratory, and the Advanced Photon Source, a DOE Office of Science User Facility operated for the DOE Office of Science by Argonne National Laboratory under Contract No. DE-AC02-06CH11357.
Z.D. acknowledges support from the Swedish Research Council (VR) under Grant No. 2015-00585, cofunded by Marie Sk\l{}odowska Curie Actions (Project INCA 600398).
\bibliography{apssamp}

\begin{thebibliography}{44}%
\makeatletter
\providecommand \@ifxundefined [1]{%
 \@ifx{#1\undefined}
}%
\providecommand \@ifnum [1]{%
 \ifnum #1\expandafter \@firstoftwo
 \else \expandafter \@secondoftwo
 \fi
}%
\providecommand \@ifx [1]{%
 \ifx #1\expandafter \@firstoftwo
 \else \expandafter \@secondoftwo
 \fi
}%
\providecommand \natexlab [1]{#1}%
\providecommand \enquote  [1]{``#1''}%
\providecommand \bibnamefont  [1]{#1}%
\providecommand \bibfnamefont [1]{#1}%
\providecommand \citenamefont [1]{#1}%
\providecommand \href@noop [0]{\@secondoftwo}%
\providecommand \href [0]{\begingroup \@sanitize@url \@href}%
\providecommand \@href[1]{\@@startlink{#1}\@@href}%
\providecommand \@@href[1]{\endgroup#1\@@endlink}%
\providecommand \@sanitize@url [0]{\catcode `\\12\catcode `\$12\catcode
  `\&12\catcode `\#12\catcode `\^12\catcode `\_12\catcode `\%12\relax}%
\providecommand \@@startlink[1]{}%
\providecommand \@@endlink[0]{}%
\providecommand \url  [0]{\begingroup\@sanitize@url \@url }%
\providecommand \@url [1]{\endgroup\@href {#1}{\urlprefix }}%
\providecommand \urlprefix  [0]{URL }%
\providecommand \Eprint [0]{\href }%
\providecommand \doibase [0]{https://doi.org/}%
\providecommand \selectlanguage [0]{\@gobble}%
\providecommand \bibinfo  [0]{\@secondoftwo}%
\providecommand \bibfield  [0]{\@secondoftwo}%
\providecommand \translation [1]{[#1]}%
\providecommand \BibitemOpen [0]{}%
\providecommand \bibitemStop [0]{}%
\providecommand \bibitemNoStop [0]{.\EOS\space}%
\providecommand \EOS [0]{\spacefactor3000\relax}%
\providecommand \BibitemShut  [1]{\csname bibitem#1\endcsname}%
\let\auto@bib@innerbib\@empty
\bibitem [{\citenamefont {Duine}(2011)}]{duine2011spintronics}%
  \BibitemOpen
  \bibfield  {author} {\bibinfo {author} {\bibfnamefont {R.}~\bibnamefont
  {Duine}},\ }\bibfield  {title} {\bibinfo {title} {Spintronics: An alternating
  alternative},\ }\href@noop {} {\bibfield  {journal} {\bibinfo  {journal}
  {Nature materials}\ }\textbf {\bibinfo {volume} {10}},\ \bibinfo {pages}
  {344} (\bibinfo {year} {2011})}\BibitemShut {NoStop}%
\bibitem [{\citenamefont {Sinova}\ and\ \citenamefont
  {{\v{Z}}uti{\'c}}(2012)}]{sinova2012new}%
  \BibitemOpen
  \bibfield  {author} {\bibinfo {author} {\bibfnamefont {J.}~\bibnamefont
  {Sinova}}\ and\ \bibinfo {author} {\bibfnamefont {I.}~\bibnamefont
  {{\v{Z}}uti{\'c}}},\ }\bibfield  {title} {\bibinfo {title} {New moves of the
  spintronics tango},\ }\href@noop {} {\bibfield  {journal} {\bibinfo
  {journal} {Nature materials}\ }\textbf {\bibinfo {volume} {11}},\ \bibinfo
  {pages} {368} (\bibinfo {year} {2012})}\BibitemShut {NoStop}%
\bibitem [{\citenamefont {Satoh}\ \emph {et~al.}(2007)\citenamefont {Satoh},
  \citenamefont {Van~Aken}, \citenamefont {Duong}, \citenamefont
  {Lottermoser},\ and\ \citenamefont {Fiebig}}]{SatohCr2O3}%
  \BibitemOpen
  \bibfield  {author} {\bibinfo {author} {\bibfnamefont {T.}~\bibnamefont
  {Satoh}}, \bibinfo {author} {\bibfnamefont {B.~B.}\ \bibnamefont {Van~Aken}},
  \bibinfo {author} {\bibfnamefont {N.~P.}\ \bibnamefont {Duong}}, \bibinfo
  {author} {\bibfnamefont {T.}~\bibnamefont {Lottermoser}},\ and\ \bibinfo
  {author} {\bibfnamefont {M.}~\bibnamefont {Fiebig}},\ }\bibfield  {title}
  {\bibinfo {title} {Ultrafast spin and lattice dynamics in antiferromagnetic
  cr2o3},\ }\href@noop {} {\bibfield  {journal} {\bibinfo  {journal} {Physical
  Review B}\ }\textbf {\bibinfo {volume} {75}},\ \bibinfo {pages} {155406}
  (\bibinfo {year} {2007})}\BibitemShut {NoStop}%
\bibitem [{\citenamefont {Tzschaschel}\ \emph {et~al.}(2017)\citenamefont
  {Tzschaschel}, \citenamefont {Otani}, \citenamefont {Iida}, \citenamefont
  {Shimura}, \citenamefont {Ueda}, \citenamefont {G{\"u}nther}, \citenamefont
  {Fiebig},\ and\ \citenamefont {Satoh}}]{SatohNiO}%
  \BibitemOpen
  \bibfield  {author} {\bibinfo {author} {\bibfnamefont {C.}~\bibnamefont
  {Tzschaschel}}, \bibinfo {author} {\bibfnamefont {K.}~\bibnamefont {Otani}},
  \bibinfo {author} {\bibfnamefont {R.}~\bibnamefont {Iida}}, \bibinfo {author}
  {\bibfnamefont {T.}~\bibnamefont {Shimura}}, \bibinfo {author} {\bibfnamefont
  {H.}~\bibnamefont {Ueda}}, \bibinfo {author} {\bibfnamefont {S.}~\bibnamefont
  {G{\"u}nther}}, \bibinfo {author} {\bibfnamefont {M.}~\bibnamefont
  {Fiebig}},\ and\ \bibinfo {author} {\bibfnamefont {T.}~\bibnamefont
  {Satoh}},\ }\bibfield  {title} {\bibinfo {title} {Ultrafast optical
  excitation of coherent magnons in antiferromagnetic nio},\ }\href@noop {}
  {\bibfield  {journal} {\bibinfo  {journal} {Physical Review B}\ }\textbf
  {\bibinfo {volume} {95}},\ \bibinfo {pages} {174407} (\bibinfo {year}
  {2017})}\BibitemShut {NoStop}%
\bibitem [{\citenamefont {Kalashnikova}\ \emph {et~al.}(2015)\citenamefont
  {Kalashnikova}, \citenamefont {Kimel},\ and\ \citenamefont
  {Pisarev}}]{PisarevReview}%
  \BibitemOpen
  \bibfield  {author} {\bibinfo {author} {\bibfnamefont {A.~M.}\ \bibnamefont
  {Kalashnikova}}, \bibinfo {author} {\bibfnamefont {A.~V.}\ \bibnamefont
  {Kimel}},\ and\ \bibinfo {author} {\bibfnamefont {R.~V.}\ \bibnamefont
  {Pisarev}},\ }\bibfield  {title} {\bibinfo {title} {Ultrafast
  opto-magnetism},\ }\href@noop {} {\bibfield  {journal} {\bibinfo  {journal}
  {Physics-Uspekhi}\ }\textbf {\bibinfo {volume} {58}},\ \bibinfo {pages} {969}
  (\bibinfo {year} {2015})}\BibitemShut {NoStop}%
\bibitem [{\citenamefont {Beaurepaire}\ \emph {et~al.}(1996)\citenamefont
  {Beaurepaire}, \citenamefont {Merle}, \citenamefont {Daunois},\ and\
  \citenamefont {Bigot}}]{Bigot_Ni}%
  \BibitemOpen
  \bibfield  {author} {\bibinfo {author} {\bibfnamefont {E.}~\bibnamefont
  {Beaurepaire}}, \bibinfo {author} {\bibfnamefont {J.-C.}\ \bibnamefont
  {Merle}}, \bibinfo {author} {\bibfnamefont {A.}~\bibnamefont {Daunois}},\
  and\ \bibinfo {author} {\bibfnamefont {J.-Y.}\ \bibnamefont {Bigot}},\
  }\bibfield  {title} {\bibinfo {title} {Ultrafast spin dynamics in
  ferromagnetic nickel},\ }\href@noop {} {\bibfield  {journal} {\bibinfo
  {journal} {Physical review letters}\ }\textbf {\bibinfo {volume} {76}},\
  \bibinfo {pages} {4250} (\bibinfo {year} {1996})}\BibitemShut {NoStop}%
\bibitem [{\citenamefont {Visnovsky}\ \emph {et~al.}(1981)\citenamefont
  {Visnovsky}, \citenamefont {Prosser}, \citenamefont {Krishnan}, \citenamefont
  {Parizek}, \citenamefont {Nitsch},\ and\ \citenamefont
  {Svobodova}}]{FiMMOKE}%
  \BibitemOpen
  \bibfield  {author} {\bibinfo {author} {\bibfnamefont {S.}~\bibnamefont
  {Visnovsky}}, \bibinfo {author} {\bibfnamefont {V.}~\bibnamefont {Prosser}},
  \bibinfo {author} {\bibfnamefont {R.}~\bibnamefont {Krishnan}}, \bibinfo
  {author} {\bibfnamefont {V.}~\bibnamefont {Parizek}}, \bibinfo {author}
  {\bibfnamefont {K.}~\bibnamefont {Nitsch}},\ and\ \bibinfo {author}
  {\bibfnamefont {L.}~\bibnamefont {Svobodova}},\ }\bibfield  {title} {\bibinfo
  {title} {Magnetooptical polar kerr effect in ferrimagnetic garnets and
  spinels},\ }\href@noop {} {\bibfield  {journal} {\bibinfo  {journal} {IEEE
  Transactions on Magnetics}\ }\textbf {\bibinfo {volume} {17}},\ \bibinfo
  {pages} {3205} (\bibinfo {year} {1981})}\BibitemShut {NoStop}%
\bibitem [{\citenamefont {McCord}(2015)}]{Kerrmicroscopy}%
  \BibitemOpen
  \bibfield  {author} {\bibinfo {author} {\bibfnamefont {J.}~\bibnamefont
  {McCord}},\ }\bibfield  {title} {\bibinfo {title} {Progress in magnetic
  domain observation by advanced magneto-optical microscopy},\ }\href@noop {}
  {\bibfield  {journal} {\bibinfo  {journal} {Journal of Physics D: Applied
  Physics}\ }\textbf {\bibinfo {volume} {48}},\ \bibinfo {pages} {333001}
  (\bibinfo {year} {2015})}\BibitemShut {NoStop}%
\bibitem [{\citenamefont {Schmool}\ \emph {et~al.}(1999)\citenamefont
  {Schmool}, \citenamefont {Keller}, \citenamefont {Guyot}, \citenamefont
  {Krishnan},\ and\ \citenamefont {Tessier}}]{cantedAFM}%
  \BibitemOpen
  \bibfield  {author} {\bibinfo {author} {\bibfnamefont {D.}~\bibnamefont
  {Schmool}}, \bibinfo {author} {\bibfnamefont {N.}~\bibnamefont {Keller}},
  \bibinfo {author} {\bibfnamefont {M.}~\bibnamefont {Guyot}}, \bibinfo
  {author} {\bibfnamefont {R.}~\bibnamefont {Krishnan}},\ and\ \bibinfo
  {author} {\bibfnamefont {M.}~\bibnamefont {Tessier}},\ }\bibfield  {title}
  {\bibinfo {title} {Magnetic and magneto-optic properties of orthoferrite thin
  films grown by pulsed-laser deposition},\ }\href@noop {} {\bibfield
  {journal} {\bibinfo  {journal} {Journal of applied physics}\ }\textbf
  {\bibinfo {volume} {86}},\ \bibinfo {pages} {5712} (\bibinfo {year}
  {1999})}\BibitemShut {NoStop}%
\bibitem [{\citenamefont {Higo}\ \emph {et~al.}(2018)\citenamefont {Higo},
  \citenamefont {Man}, \citenamefont {Gopman}, \citenamefont {Wu},
  \citenamefont {Koretsune}, \citenamefont {van’t Erve}, \citenamefont
  {Kabanov}, \citenamefont {Rees}, \citenamefont {Li}, \citenamefont {Suzuki}
  \emph {et~al.}}]{Mn3Sn}%
  \BibitemOpen
  \bibfield  {author} {\bibinfo {author} {\bibfnamefont {T.}~\bibnamefont
  {Higo}}, \bibinfo {author} {\bibfnamefont {H.}~\bibnamefont {Man}}, \bibinfo
  {author} {\bibfnamefont {D.~B.}\ \bibnamefont {Gopman}}, \bibinfo {author}
  {\bibfnamefont {L.}~\bibnamefont {Wu}}, \bibinfo {author} {\bibfnamefont
  {T.}~\bibnamefont {Koretsune}}, \bibinfo {author} {\bibfnamefont {O.~M.}\
  \bibnamefont {van’t Erve}}, \bibinfo {author} {\bibfnamefont {Y.~P.}\
  \bibnamefont {Kabanov}}, \bibinfo {author} {\bibfnamefont {D.}~\bibnamefont
  {Rees}}, \bibinfo {author} {\bibfnamefont {Y.}~\bibnamefont {Li}}, \bibinfo
  {author} {\bibfnamefont {M.-T.}\ \bibnamefont {Suzuki}}, \emph {et~al.},\
  }\bibfield  {title} {\bibinfo {title} {Large magneto-optical kerr effect and
  imaging of magnetic octupole domains in an antiferromagnetic metal},\
  }\href@noop {} {\bibfield  {journal} {\bibinfo  {journal} {Nature photonics}\
  }\textbf {\bibinfo {volume} {12}},\ \bibinfo {pages} {73} (\bibinfo {year}
  {2018})}\BibitemShut {NoStop}%
\bibitem [{\citenamefont {Nemec}\ \emph {et~al.}(2017)\citenamefont {Nemec},
  \citenamefont {Fiebig}, \citenamefont {Kampfrath},\ and\ \citenamefont
  {Kimel}}]{Kimelreview}%
  \BibitemOpen
  \bibfield  {author} {\bibinfo {author} {\bibfnamefont {P.}~\bibnamefont
  {Nemec}}, \bibinfo {author} {\bibfnamefont {M.}~\bibnamefont {Fiebig}},
  \bibinfo {author} {\bibfnamefont {T.}~\bibnamefont {Kampfrath}},\ and\
  \bibinfo {author} {\bibfnamefont {A.}~\bibnamefont {Kimel}},\ }\bibfield
  {title} {\bibinfo {title} {Antiferromagnetic opto-spintronics: Part of a
  collection of reviews on antiferromagnetic spintronics},\ }\href@noop {}
  {\bibfield  {journal} {\bibinfo  {journal} {arXiv preprint arXiv:1705.10600}\
  } (\bibinfo {year} {2017})}\BibitemShut {NoStop}%
\bibitem [{\citenamefont {Muir}\ and\ \citenamefont
  {Str{\"o}m-Olsen}(1971)}]{MuirCr}%
  \BibitemOpen
  \bibfield  {author} {\bibinfo {author} {\bibfnamefont {W.}~\bibnamefont
  {Muir}}\ and\ \bibinfo {author} {\bibfnamefont {J.}~\bibnamefont
  {Str{\"o}m-Olsen}},\ }\bibfield  {title} {\bibinfo {title} {Electrical
  resistance of single-crystal single-domain chromium from 77 to 325 k},\
  }\href@noop {} {\bibfield  {journal} {\bibinfo  {journal} {Physical Review
  B}\ }\textbf {\bibinfo {volume} {4}},\ \bibinfo {pages} {988} (\bibinfo
  {year} {1971})}\BibitemShut {NoStop}%
\bibitem [{\citenamefont {Jungwirth}\ \emph {et~al.}(2016)\citenamefont
  {Jungwirth}, \citenamefont {Marti}, \citenamefont {Wadley},\ and\
  \citenamefont {Wunderlich}}]{jungwirth2016antiferromagnetic}%
  \BibitemOpen
  \bibfield  {author} {\bibinfo {author} {\bibfnamefont {T.}~\bibnamefont
  {Jungwirth}}, \bibinfo {author} {\bibfnamefont {X.}~\bibnamefont {Marti}},
  \bibinfo {author} {\bibfnamefont {P.}~\bibnamefont {Wadley}},\ and\ \bibinfo
  {author} {\bibfnamefont {J.}~\bibnamefont {Wunderlich}},\ }\bibfield  {title}
  {\bibinfo {title} {Antiferromagnetic spintronics},\ }\href@noop {} {\bibfield
   {journal} {\bibinfo  {journal} {Nature nanotechnology}\ }\textbf {\bibinfo
  {volume} {11}},\ \bibinfo {pages} {231} (\bibinfo {year} {2016})}\BibitemShut
  {NoStop}%
\bibitem [{\citenamefont {Bodnar}\ \emph {et~al.}(2018)\citenamefont {Bodnar},
  \citenamefont {{\v{S}}mejkal}, \citenamefont {Turek}, \citenamefont
  {Jungwirth}, \citenamefont {Gomonay}, \citenamefont {Sinova}, \citenamefont
  {Sapozhnik}, \citenamefont {Elmers}, \citenamefont {Kl{\"a}ui},\ and\
  \citenamefont {Jourdan}}]{Mn2Au}%
  \BibitemOpen
  \bibfield  {author} {\bibinfo {author} {\bibfnamefont {S.~Y.}\ \bibnamefont
  {Bodnar}}, \bibinfo {author} {\bibfnamefont {L.}~\bibnamefont
  {{\v{S}}mejkal}}, \bibinfo {author} {\bibfnamefont {I.}~\bibnamefont
  {Turek}}, \bibinfo {author} {\bibfnamefont {T.}~\bibnamefont {Jungwirth}},
  \bibinfo {author} {\bibfnamefont {O.}~\bibnamefont {Gomonay}}, \bibinfo
  {author} {\bibfnamefont {J.}~\bibnamefont {Sinova}}, \bibinfo {author}
  {\bibfnamefont {A.}~\bibnamefont {Sapozhnik}}, \bibinfo {author}
  {\bibfnamefont {H.-J.}\ \bibnamefont {Elmers}}, \bibinfo {author}
  {\bibfnamefont {M.}~\bibnamefont {Kl{\"a}ui}},\ and\ \bibinfo {author}
  {\bibfnamefont {M.}~\bibnamefont {Jourdan}},\ }\bibfield  {title} {\bibinfo
  {title} {Writing and reading antiferromagnetic mn 2 au by n{\'e}el spin-orbit
  torques and large anisotropic magnetoresistance},\ }\href@noop {} {\bibfield
  {journal} {\bibinfo  {journal} {Nature Communications}\ }\textbf {\bibinfo
  {volume} {9}},\ \bibinfo {pages} {348} (\bibinfo {year} {2018})}\BibitemShut
  {NoStop}%
\bibitem [{\citenamefont {Dhesi}\ \emph {et~al.}(2002)\citenamefont {Dhesi},
  \citenamefont {van~der Laan},\ and\ \citenamefont {Dudzik}}]{XMLD_APL}%
  \BibitemOpen
  \bibfield  {author} {\bibinfo {author} {\bibfnamefont {S.}~\bibnamefont
  {Dhesi}}, \bibinfo {author} {\bibfnamefont {G.}~\bibnamefont {van~der
  Laan}},\ and\ \bibinfo {author} {\bibfnamefont {E.}~\bibnamefont {Dudzik}},\
  }\bibfield  {title} {\bibinfo {title} {Determining element-specific
  magnetocrystalline anisotropies using x-ray magnetic linear dichroism},\
  }\href@noop {} {\bibfield  {journal} {\bibinfo  {journal} {Applied physics
  letters}\ }\textbf {\bibinfo {volume} {80}},\ \bibinfo {pages} {1613}
  (\bibinfo {year} {2002})}\BibitemShut {NoStop}%
\bibitem [{\citenamefont {Ferr{\'e}}\ and\ \citenamefont
  {Gehring}(1984)}]{ferre1984linear}%
  \BibitemOpen
  \bibfield  {author} {\bibinfo {author} {\bibfnamefont {J.}~\bibnamefont
  {Ferr{\'e}}}\ and\ \bibinfo {author} {\bibfnamefont {G.}~\bibnamefont
  {Gehring}},\ }\bibfield  {title} {\bibinfo {title} {Linear optical
  birefringence of magnetic crystals},\ }\href@noop {} {\bibfield  {journal}
  {\bibinfo  {journal} {Reports on Progress in Physics}\ }\textbf {\bibinfo
  {volume} {47}},\ \bibinfo {pages} {513} (\bibinfo {year} {1984})}\BibitemShut
  {NoStop}%
\bibitem [{\citenamefont {Saidl}\ \emph {et~al.}(2017)\citenamefont {Saidl},
  \citenamefont {N{\v{e}}mec}, \citenamefont {Wadley}, \citenamefont {Hills},
  \citenamefont {Campion}, \citenamefont {Nov{\'a}k}, \citenamefont {Edmonds},
  \citenamefont {Maccherozzi}, \citenamefont {Dhesi}, \citenamefont {Gallagher}
  \emph {et~al.}}]{CuMnAs}%
  \BibitemOpen
  \bibfield  {author} {\bibinfo {author} {\bibfnamefont {V.}~\bibnamefont
  {Saidl}}, \bibinfo {author} {\bibfnamefont {P.}~\bibnamefont {N{\v{e}}mec}},
  \bibinfo {author} {\bibfnamefont {P.}~\bibnamefont {Wadley}}, \bibinfo
  {author} {\bibfnamefont {V.}~\bibnamefont {Hills}}, \bibinfo {author}
  {\bibfnamefont {R.}~\bibnamefont {Campion}}, \bibinfo {author} {\bibfnamefont
  {V.}~\bibnamefont {Nov{\'a}k}}, \bibinfo {author} {\bibfnamefont
  {K.}~\bibnamefont {Edmonds}}, \bibinfo {author} {\bibfnamefont
  {F.}~\bibnamefont {Maccherozzi}}, \bibinfo {author} {\bibfnamefont
  {S.}~\bibnamefont {Dhesi}}, \bibinfo {author} {\bibfnamefont
  {B.}~\bibnamefont {Gallagher}}, \emph {et~al.},\ }\bibfield  {title}
  {\bibinfo {title} {Optical determination of the n{\'e}el vector in a cumnas
  thin-film antiferromagnet},\ }\href@noop {} {\bibfield  {journal} {\bibinfo
  {journal} {Nature Photonics}\ }\textbf {\bibinfo {volume} {11}},\ \bibinfo
  {pages} {91} (\bibinfo {year} {2017})}\BibitemShut {NoStop}%
\bibitem [{\citenamefont {Katsuraki}\ and\ \citenamefont
  {Achiwa}(1966)}]{Fe2Asstructure}%
  \BibitemOpen
  \bibfield  {author} {\bibinfo {author} {\bibfnamefont {H.}~\bibnamefont
  {Katsuraki}}\ and\ \bibinfo {author} {\bibfnamefont {N.}~\bibnamefont
  {Achiwa}},\ }\bibfield  {title} {\bibinfo {title} {The magnetic structure of
  fe2as},\ }\href@noop {} {\bibfield  {journal} {\bibinfo  {journal} {Journal
  of the Physical Society of Japan}\ }\textbf {\bibinfo {volume} {21}},\
  \bibinfo {pages} {2238} (\bibinfo {year} {1966})}\BibitemShut {NoStop}%
\bibitem [{\citenamefont {Le~Gall}\ \emph {et~al.}(1971)\citenamefont
  {Le~Gall}, \citenamefont {Vien},\ and\ \citenamefont {Desormiere}}]{LeGall}%
  \BibitemOpen
  \bibfield  {author} {\bibinfo {author} {\bibfnamefont {H.}~\bibnamefont
  {Le~Gall}}, \bibinfo {author} {\bibfnamefont {T.~K.}\ \bibnamefont {Vien}},\
  and\ \bibinfo {author} {\bibfnamefont {B.}~\bibnamefont {Desormiere}},\
  }\bibfield  {title} {\bibinfo {title} {Theory of the elastic and inelastic
  scattering of light by magnetic crystals. ii. second-order processes},\
  }\href@noop {} {\bibfield  {journal} {\bibinfo  {journal} {physica status
  solidi (b)}\ }\textbf {\bibinfo {volume} {47}},\ \bibinfo {pages} {591}
  (\bibinfo {year} {1971})}\BibitemShut {NoStop}%
\bibitem [{\citenamefont {Tesa{\v{r}}ov{\'a}}\ \emph
  {et~al.}(2012)\citenamefont {Tesa{\v{r}}ov{\'a}}, \citenamefont
  {N{\v{e}}mec}, \citenamefont {Rozkotov{\'a}}, \citenamefont {{\v{S}}ubrt},
  \citenamefont {Reichlov{\'a}}, \citenamefont {Butkovi{\v{c}}ov{\'a}},
  \citenamefont {Troj{\'a}nek}, \citenamefont {Mal{\`y}}, \citenamefont
  {Nov{\'a}k},\ and\ \citenamefont {Jungwirth}}]{tesavrova3Dtrajectory}%
  \BibitemOpen
  \bibfield  {author} {\bibinfo {author} {\bibfnamefont {N.}~\bibnamefont
  {Tesa{\v{r}}ov{\'a}}}, \bibinfo {author} {\bibfnamefont {P.}~\bibnamefont
  {N{\v{e}}mec}}, \bibinfo {author} {\bibfnamefont {E.}~\bibnamefont
  {Rozkotov{\'a}}}, \bibinfo {author} {\bibfnamefont {J.}~\bibnamefont
  {{\v{S}}ubrt}}, \bibinfo {author} {\bibfnamefont {H.}~\bibnamefont
  {Reichlov{\'a}}}, \bibinfo {author} {\bibfnamefont {D.}~\bibnamefont
  {Butkovi{\v{c}}ov{\'a}}}, \bibinfo {author} {\bibfnamefont {F.}~\bibnamefont
  {Troj{\'a}nek}}, \bibinfo {author} {\bibfnamefont {P.}~\bibnamefont
  {Mal{\`y}}}, \bibinfo {author} {\bibfnamefont {V.}~\bibnamefont
  {Nov{\'a}k}},\ and\ \bibinfo {author} {\bibfnamefont {T.}~\bibnamefont
  {Jungwirth}},\ }\bibfield  {title} {\bibinfo {title} {Direct measurement of
  the three-dimensional magnetization vector trajectory in gamnas by a
  magneto-optical pump-and-probe method},\ }\href@noop {} {\bibfield  {journal}
  {\bibinfo  {journal} {Applied Physics Letters}\ }\textbf {\bibinfo {volume}
  {100}},\ \bibinfo {pages} {102403} (\bibinfo {year} {2012})}\BibitemShut
  {NoStop}%
\bibitem [{\citenamefont {Kresse}\ and\ \citenamefont
  {Furthm{\"u}ller}(1996)}]{Kresse:1996}%
  \BibitemOpen
  \bibfield  {author} {\bibinfo {author} {\bibfnamefont {G.}~\bibnamefont
  {Kresse}}\ and\ \bibinfo {author} {\bibfnamefont {J.}~\bibnamefont
  {Furthm{\"u}ller}},\ }\bibfield  {title} {\bibinfo {title} {Efficient
  iterative schemes for ab initio total-energy calculations using a plane-wave
  basis set},\ }\href {https://doi.org/10.1103/PhysRevB.54.11169} {\bibfield
  {journal} {\bibinfo  {journal} {Phys. Rev. B}\ }\textbf {\bibinfo {volume}
  {54}},\ \bibinfo {pages} {11169} (\bibinfo {year} {1996})}\BibitemShut
  {NoStop}%
\bibitem [{\citenamefont {Kresse}\ and\ \citenamefont
  {Joubert}(1999)}]{Kresse:1999}%
  \BibitemOpen
  \bibfield  {author} {\bibinfo {author} {\bibfnamefont {G.}~\bibnamefont
  {Kresse}}\ and\ \bibinfo {author} {\bibfnamefont {D.}~\bibnamefont
  {Joubert}},\ }\bibfield  {title} {\bibinfo {title} {From ultrasoft
  pseudopotentials to the projector augmented-wave method},\ }\href
  {https://doi.org/10.1103/PhysRevB.59.1758} {\bibfield  {journal} {\bibinfo
  {journal} {Phys. Rev. B}\ }\textbf {\bibinfo {volume} {59}},\ \bibinfo
  {pages} {1758} (\bibinfo {year} {1999})}\BibitemShut {NoStop}%
\bibitem [{\citenamefont {Gajdo\v{s}}\ \emph {et~al.}(2006)\citenamefont
  {Gajdo\v{s}}, \citenamefont {Hummer}, \citenamefont {Kresse}, \citenamefont
  {Furthm\"uller},\ and\ \citenamefont {Bechstedt}}]{Gajdos:2006}%
  \BibitemOpen
  \bibfield  {author} {\bibinfo {author} {\bibfnamefont {M.}~\bibnamefont
  {Gajdo\v{s}}}, \bibinfo {author} {\bibfnamefont {K.}~\bibnamefont {Hummer}},
  \bibinfo {author} {\bibfnamefont {G.}~\bibnamefont {Kresse}}, \bibinfo
  {author} {\bibfnamefont {J.}~\bibnamefont {Furthm\"uller}},\ and\ \bibinfo
  {author} {\bibfnamefont {F.}~\bibnamefont {Bechstedt}},\ }\bibfield  {title}
  {\bibinfo {title} {Linear optical properties in the projector-augmented wave
  methodology},\ }\href {https://doi.org/10.1103/PhysRevB.73.045112} {\bibfield
   {journal} {\bibinfo  {journal} {Phys. Rev. B}\ }\textbf {\bibinfo {volume}
  {73}},\ \bibinfo {pages} {045112} (\bibinfo {year} {2006})}\BibitemShut
  {NoStop}%
\bibitem [{\citenamefont {Perdew}\ \emph {et~al.}(1996)\citenamefont {Perdew},
  \citenamefont {Burke},\ and\ \citenamefont {Ernzerhof}}]{Perdew:1997}%
  \BibitemOpen
  \bibfield  {author} {\bibinfo {author} {\bibfnamefont {J.~P.}\ \bibnamefont
  {Perdew}}, \bibinfo {author} {\bibfnamefont {K.}~\bibnamefont {Burke}},\ and\
  \bibinfo {author} {\bibfnamefont {M.}~\bibnamefont {Ernzerhof}},\ }\bibfield
  {title} {\bibinfo {title} {Generalized gradient approximation made simple},\
  }\href {https://doi.org/10.1103/PhysRevLett.77.3865} {\bibfield  {journal}
  {\bibinfo  {journal} {Phys. Rev. Lett.}\ }\textbf {\bibinfo {volume} {77}},\
  \bibinfo {pages} {3865} (\bibinfo {year} {1996})}\BibitemShut {NoStop}%
\bibitem [{\citenamefont {Bl\"ochl}(1994)}]{Blochl:1994}%
  \BibitemOpen
  \bibfield  {author} {\bibinfo {author} {\bibfnamefont {P.~E.}\ \bibnamefont
  {Bl\"ochl}},\ }\bibfield  {title} {\bibinfo {title} {Projector augmented-wave
  method},\ }\href {https://doi.org/10.1103/PhysRevB.50.17953} {\bibfield
  {journal} {\bibinfo  {journal} {Phys. Rev. B}\ }\textbf {\bibinfo {volume}
  {50}},\ \bibinfo {pages} {17953} (\bibinfo {year} {1994})}\BibitemShut
  {NoStop}%
\bibitem [{\citenamefont {Monkhorst}\ and\ \citenamefont
  {Pack}(1976)}]{Monkhorst:1976}%
  \BibitemOpen
  \bibfield  {author} {\bibinfo {author} {\bibfnamefont {H.~J.}\ \bibnamefont
  {Monkhorst}}\ and\ \bibinfo {author} {\bibfnamefont {J.~D.}\ \bibnamefont
  {Pack}},\ }\bibfield  {title} {\bibinfo {title} {Special points for
  brillouin-zone integrations},\ }\href
  {https://doi.org/10.1103/PhysRevB.13.5188} {\bibfield  {journal} {\bibinfo
  {journal} {Phys. Rev. B}\ }\textbf {\bibinfo {volume} {13}},\ \bibinfo
  {pages} {5188} (\bibinfo {year} {1976})}\BibitemShut {NoStop}%
\bibitem [{\citenamefont {Togo}\ and\ \citenamefont
  {Tanaka}(2015)}]{Phonopy:2015}%
  \BibitemOpen
  \bibfield  {author} {\bibinfo {author} {\bibfnamefont {A.}~\bibnamefont
  {Togo}}\ and\ \bibinfo {author} {\bibfnamefont {I.}~\bibnamefont {Tanaka}},\
  }\bibfield  {title} {\bibinfo {title} {First principles phonon calculations
  in materials science},\ }\href
  {https://doi.org/10.1016/j.scriptamat.2015.07.021} {\bibfield  {journal}
  {\bibinfo  {journal} {Scr. Mater.}\ }\textbf {\bibinfo {volume} {108}},\
  \bibinfo {pages} {1} (\bibinfo {year} {2015})}\BibitemShut {NoStop}%
\bibitem [{\citenamefont {Corliss}\ \emph {et~al.}(1982)\citenamefont
  {Corliss}, \citenamefont {Hastings}, \citenamefont {Kunnmann}, \citenamefont
  {Begum}, \citenamefont {Collins}, \citenamefont {Gurewitz},\ and\
  \citenamefont {Mukamel}}]{phasediagramFe2As}%
  \BibitemOpen
  \bibfield  {author} {\bibinfo {author} {\bibfnamefont {L.}~\bibnamefont
  {Corliss}}, \bibinfo {author} {\bibfnamefont {J.}~\bibnamefont {Hastings}},
  \bibinfo {author} {\bibfnamefont {W.}~\bibnamefont {Kunnmann}}, \bibinfo
  {author} {\bibfnamefont {R.}~\bibnamefont {Begum}}, \bibinfo {author}
  {\bibfnamefont {M.}~\bibnamefont {Collins}}, \bibinfo {author} {\bibfnamefont
  {E.}~\bibnamefont {Gurewitz}},\ and\ \bibinfo {author} {\bibfnamefont
  {D.}~\bibnamefont {Mukamel}},\ }\bibfield  {title} {\bibinfo {title}
  {Magnetic phase diagram and critical behavior of fe 2 as},\ }\href@noop {}
  {\bibfield  {journal} {\bibinfo  {journal} {Physical Review B}\ }\textbf
  {\bibinfo {volume} {25}},\ \bibinfo {pages} {245} (\bibinfo {year}
  {1982})}\BibitemShut {NoStop}%
\bibitem [{\citenamefont {Achiwa}\ \emph {et~al.}(1967)\citenamefont {Achiwa},
  \citenamefont {Yano}, \citenamefont {Yuzuri},\ and\ \citenamefont
  {Takaki}}]{AchiwaTorque}%
  \BibitemOpen
  \bibfield  {author} {\bibinfo {author} {\bibfnamefont {N.}~\bibnamefont
  {Achiwa}}, \bibinfo {author} {\bibfnamefont {S.}~\bibnamefont {Yano}},
  \bibinfo {author} {\bibfnamefont {M.}~\bibnamefont {Yuzuri}},\ and\ \bibinfo
  {author} {\bibfnamefont {H.}~\bibnamefont {Takaki}},\ }\bibfield  {title}
  {\bibinfo {title} {Magnetic anisotropy in the c-plane of fe2as},\ }\href@noop
  {} {\bibfield  {journal} {\bibinfo  {journal} {Journal of the Physical
  Society of Japan}\ }\textbf {\bibinfo {volume} {22}},\ \bibinfo {pages} {156}
  (\bibinfo {year} {1967})}\BibitemShut {NoStop}%
\bibitem [{\citenamefont {Zocco}\ \emph {et~al.}(2012)\citenamefont {Zocco},
  \citenamefont {T{\"u}t{\"u}n}, \citenamefont {Hamlin}, \citenamefont
  {Jeffries}, \citenamefont {Weir}, \citenamefont {Vohra},\ and\ \citenamefont
  {Maple}}]{zocco2012high}%
  \BibitemOpen
  \bibfield  {author} {\bibinfo {author} {\bibfnamefont {D.~A.}\ \bibnamefont
  {Zocco}}, \bibinfo {author} {\bibfnamefont {D.~Y.}\ \bibnamefont
  {T{\"u}t{\"u}n}}, \bibinfo {author} {\bibfnamefont {J.~J.}\ \bibnamefont
  {Hamlin}}, \bibinfo {author} {\bibfnamefont {J.~R.}\ \bibnamefont
  {Jeffries}}, \bibinfo {author} {\bibfnamefont {S.~T.}\ \bibnamefont {Weir}},
  \bibinfo {author} {\bibfnamefont {Y.~K.}\ \bibnamefont {Vohra}},\ and\
  \bibinfo {author} {\bibfnamefont {M.~B.}\ \bibnamefont {Maple}},\ }\bibfield
  {title} {\bibinfo {title} {High pressure transport studies of the lifeas
  analogs cufete2 and fe2as},\ }\href@noop {} {\bibfield  {journal} {\bibinfo
  {journal} {Superconductor Science and Technology}\ }\textbf {\bibinfo
  {volume} {25}},\ \bibinfo {pages} {084018} (\bibinfo {year}
  {2012})}\BibitemShut {NoStop}%
\bibitem [{\citenamefont {Mermin}(1965)}]{Mermin:1965}%
  \BibitemOpen
  \bibfield  {author} {\bibinfo {author} {\bibfnamefont {N.~D.}\ \bibnamefont
  {Mermin}},\ }\bibfield  {title} {\bibinfo {title} {Thermal properties of the
  inhomogeneous electron gas},\ }\href
  {https://doi.org/10.1103/PhysRev.137.A1441} {\bibfield  {journal} {\bibinfo
  {journal} {Phys. Rev.}\ }\textbf {\bibinfo {volume} {137}},\ \bibinfo {pages}
  {A1441} (\bibinfo {year} {1965})}\BibitemShut {NoStop}%
\bibitem [{\citenamefont {Cahill}(2004)}]{RSI_cahill}%
  \BibitemOpen
  \bibfield  {author} {\bibinfo {author} {\bibfnamefont {D.~G.}\ \bibnamefont
  {Cahill}},\ }\bibfield  {title} {\bibinfo {title} {Analysis of heat flow in
  layered structures for time-domain thermoreflectance},\ }\href@noop {}
  {\bibfield  {journal} {\bibinfo  {journal} {Review of scientific
  instruments}\ }\textbf {\bibinfo {volume} {75}},\ \bibinfo {pages} {5119}
  (\bibinfo {year} {2004})}\BibitemShut {NoStop}%
\bibitem [{\citenamefont {Liu}\ \emph {et~al.}(2014)\citenamefont {Liu},
  \citenamefont {Choi},\ and\ \citenamefont {Cahill}}]{liu2014measurement}%
  \BibitemOpen
  \bibfield  {author} {\bibinfo {author} {\bibfnamefont {J.}~\bibnamefont
  {Liu}}, \bibinfo {author} {\bibfnamefont {G.-M.}\ \bibnamefont {Choi}},\ and\
  \bibinfo {author} {\bibfnamefont {D.~G.}\ \bibnamefont {Cahill}},\ }\bibfield
   {title} {\bibinfo {title} {Measurement of the anisotropic thermal
  conductivity of molybdenum disulfide by the time-resolved magneto-optic kerr
  effect},\ }\href@noop {} {\bibfield  {journal} {\bibinfo  {journal} {Journal
  of Applied Physics}\ }\textbf {\bibinfo {volume} {116}},\ \bibinfo {pages}
  {233107} (\bibinfo {year} {2014})}\BibitemShut {NoStop}%
\bibitem [{\citenamefont {Kimling}\ \emph {et~al.}(2017)\citenamefont
  {Kimling}, \citenamefont {Philippi-Kobs}, \citenamefont {Jacobsohn},
  \citenamefont {Oepen},\ and\ \citenamefont {Cahill}}]{kimling2017thermal}%
  \BibitemOpen
  \bibfield  {author} {\bibinfo {author} {\bibfnamefont {J.}~\bibnamefont
  {Kimling}}, \bibinfo {author} {\bibfnamefont {A.}~\bibnamefont
  {Philippi-Kobs}}, \bibinfo {author} {\bibfnamefont {J.}~\bibnamefont
  {Jacobsohn}}, \bibinfo {author} {\bibfnamefont {H.~P.}\ \bibnamefont
  {Oepen}},\ and\ \bibinfo {author} {\bibfnamefont {D.~G.}\ \bibnamefont
  {Cahill}},\ }\bibfield  {title} {\bibinfo {title} {Thermal conductance of
  interfaces with amorphous sio 2 measured by time-resolved magneto-optic
  kerr-effect thermometry},\ }\href@noop {} {\bibfield  {journal} {\bibinfo
  {journal} {Physical Review B}\ }\textbf {\bibinfo {volume} {95}},\ \bibinfo
  {pages} {184305} (\bibinfo {year} {2017})}\BibitemShut {NoStop}%
\bibitem [{\citenamefont {Kimling}\ \emph {et~al.}(2014)\citenamefont
  {Kimling}, \citenamefont {Kimling}, \citenamefont {Wilson}, \citenamefont
  {Hebler}, \citenamefont {Albrecht},\ and\ \citenamefont
  {Cahill}}]{Kimling2014}%
  \BibitemOpen
  \bibfield  {author} {\bibinfo {author} {\bibfnamefont {J.}~\bibnamefont
  {Kimling}}, \bibinfo {author} {\bibfnamefont {J.}~\bibnamefont {Kimling}},
  \bibinfo {author} {\bibfnamefont {R.}~\bibnamefont {Wilson}}, \bibinfo
  {author} {\bibfnamefont {B.}~\bibnamefont {Hebler}}, \bibinfo {author}
  {\bibfnamefont {M.}~\bibnamefont {Albrecht}},\ and\ \bibinfo {author}
  {\bibfnamefont {D.~G.}\ \bibnamefont {Cahill}},\ }\bibfield  {title}
  {\bibinfo {title} {Ultrafast demagnetization of fept: Cu thin films and the
  role of magnetic heat capacity},\ }\href@noop {} {\bibfield  {journal}
  {\bibinfo  {journal} {Physical Review B}\ }\textbf {\bibinfo {volume} {90}},\
  \bibinfo {pages} {224408} (\bibinfo {year} {2014})}\BibitemShut {NoStop}%
\bibitem [{\citenamefont {Grzybowski}\ \emph {et~al.}(2017)\citenamefont
  {Grzybowski}, \citenamefont {Wadley}, \citenamefont {Edmonds}, \citenamefont
  {Beardsley}, \citenamefont {Hills}, \citenamefont {Campion}, \citenamefont
  {Gallagher}, \citenamefont {Chauhan}, \citenamefont {Novak}, \citenamefont
  {Jungwirth} \emph {et~al.}}]{grzybowski2017imaging}%
  \BibitemOpen
  \bibfield  {author} {\bibinfo {author} {\bibfnamefont {M.}~\bibnamefont
  {Grzybowski}}, \bibinfo {author} {\bibfnamefont {P.}~\bibnamefont {Wadley}},
  \bibinfo {author} {\bibfnamefont {K.}~\bibnamefont {Edmonds}}, \bibinfo
  {author} {\bibfnamefont {R.}~\bibnamefont {Beardsley}}, \bibinfo {author}
  {\bibfnamefont {V.}~\bibnamefont {Hills}}, \bibinfo {author} {\bibfnamefont
  {R.}~\bibnamefont {Campion}}, \bibinfo {author} {\bibfnamefont
  {B.}~\bibnamefont {Gallagher}}, \bibinfo {author} {\bibfnamefont {J.~S.}\
  \bibnamefont {Chauhan}}, \bibinfo {author} {\bibfnamefont {V.}~\bibnamefont
  {Novak}}, \bibinfo {author} {\bibfnamefont {T.}~\bibnamefont {Jungwirth}},
  \emph {et~al.},\ }\bibfield  {title} {\bibinfo {title} {Imaging
  current-induced switching of antiferromagnetic domains in cumnas},\
  }\href@noop {} {\bibfield  {journal} {\bibinfo  {journal} {Physical review
  letters}\ }\textbf {\bibinfo {volume} {118}},\ \bibinfo {pages} {057701}
  (\bibinfo {year} {2017})}\BibitemShut {NoStop}%
\bibitem [{\citenamefont {Sapozhnik}\ \emph {et~al.}(2018)\citenamefont
  {Sapozhnik}, \citenamefont {Filianina}, \citenamefont {Bodnar}, \citenamefont
  {Lamirand}, \citenamefont {Mawass}, \citenamefont {Skourski}, \citenamefont
  {Elmers}, \citenamefont {Zabel}, \citenamefont {Kl{\"a}ui},\ and\
  \citenamefont {Jourdan}}]{sapozhnik2018direct}%
  \BibitemOpen
  \bibfield  {author} {\bibinfo {author} {\bibfnamefont {A.}~\bibnamefont
  {Sapozhnik}}, \bibinfo {author} {\bibfnamefont {M.}~\bibnamefont
  {Filianina}}, \bibinfo {author} {\bibfnamefont {S.~Y.}\ \bibnamefont
  {Bodnar}}, \bibinfo {author} {\bibfnamefont {A.}~\bibnamefont {Lamirand}},
  \bibinfo {author} {\bibfnamefont {M.-A.}\ \bibnamefont {Mawass}}, \bibinfo
  {author} {\bibfnamefont {Y.}~\bibnamefont {Skourski}}, \bibinfo {author}
  {\bibfnamefont {H.-J.}\ \bibnamefont {Elmers}}, \bibinfo {author}
  {\bibfnamefont {H.}~\bibnamefont {Zabel}}, \bibinfo {author} {\bibfnamefont
  {M.}~\bibnamefont {Kl{\"a}ui}},\ and\ \bibinfo {author} {\bibfnamefont
  {M.}~\bibnamefont {Jourdan}},\ }\bibfield  {title} {\bibinfo {title} {Direct
  imaging of antiferromagnetic domains in mn 2 au manipulated by high magnetic
  fields},\ }\href@noop {} {\bibfield  {journal} {\bibinfo  {journal} {Physical
  Review B}\ }\textbf {\bibinfo {volume} {97}},\ \bibinfo {pages} {134429}
  (\bibinfo {year} {2018})}\BibitemShut {NoStop}%
\bibitem [{\citenamefont {Pisarev}\ \emph {et~al.}(1971)\citenamefont
  {Pisarev}, \citenamefont {Sinii}, \citenamefont {Kolpakova},\ and\
  \citenamefont {Yakovlev}}]{pisarev1971magnetic}%
  \BibitemOpen
  \bibfield  {author} {\bibinfo {author} {\bibfnamefont {R.}~\bibnamefont
  {Pisarev}}, \bibinfo {author} {\bibfnamefont {I.}~\bibnamefont {Sinii}},
  \bibinfo {author} {\bibfnamefont {N.}~\bibnamefont {Kolpakova}},\ and\
  \bibinfo {author} {\bibfnamefont {Y.~M.}\ \bibnamefont {Yakovlev}},\
  }\bibfield  {title} {\bibinfo {title} {Magnetic birefringence of light in
  iron garnets},\ }\href@noop {} {\bibfield  {journal} {\bibinfo  {journal}
  {Sov. Phys. JETP}\ }\textbf {\bibinfo {volume} {33}},\ \bibinfo {pages}
  {1175} (\bibinfo {year} {1971})}\BibitemShut {NoStop}%
\bibitem [{\citenamefont {Jauch}(1991)}]{jauch1991structural}%
  \BibitemOpen
  \bibfield  {author} {\bibinfo {author} {\bibfnamefont {W.}~\bibnamefont
  {Jauch}},\ }\bibfield  {title} {\bibinfo {title} {Structural origin of
  magnetic birefringence in rutile-type antiferromagnets},\ }\href@noop {}
  {\bibfield  {journal} {\bibinfo  {journal} {Physical Review B}\ }\textbf
  {\bibinfo {volume} {44}},\ \bibinfo {pages} {6864} (\bibinfo {year}
  {1991})}\BibitemShut {NoStop}%
\bibitem [{\citenamefont {Weber}\ \emph {et~al.}(2001)\citenamefont {Weber},
  \citenamefont {Bethke},\ and\ \citenamefont {Hillebrecht}}]{NiOimaging}%
  \BibitemOpen
  \bibfield  {author} {\bibinfo {author} {\bibfnamefont {N.}~\bibnamefont
  {Weber}}, \bibinfo {author} {\bibfnamefont {C.}~\bibnamefont {Bethke}},\ and\
  \bibinfo {author} {\bibfnamefont {F.}~\bibnamefont {Hillebrecht}},\
  }\bibfield  {title} {\bibinfo {title} {Imaging of antiferromagnetic domains
  at the nio (1 0 0) surface by linear dichroism in near uv photoemission
  microscopy},\ }\href@noop {} {\bibfield  {journal} {\bibinfo  {journal}
  {Journal of magnetism and magnetic materials}\ }\textbf {\bibinfo {volume}
  {226}},\ \bibinfo {pages} {1573} (\bibinfo {year} {2001})}\BibitemShut
  {NoStop}%
\bibitem [{\citenamefont {Jahn}\ and\ \citenamefont
  {Dachs}(1971)}]{jahn1971change}%
  \BibitemOpen
  \bibfield  {author} {\bibinfo {author} {\bibfnamefont {I.}~\bibnamefont
  {Jahn}}\ and\ \bibinfo {author} {\bibfnamefont {H.}~\bibnamefont {Dachs}},\
  }\bibfield  {title} {\bibinfo {title} {Change of the optical birefringence
  associated with the antiferromagnetic ordering of mnf2, fef2, cof2, and
  nif2},\ }\href@noop {} {\bibfield  {journal} {\bibinfo  {journal} {Solid
  State Communications}\ }\textbf {\bibinfo {volume} {9}},\ \bibinfo {pages}
  {1617} (\bibinfo {year} {1971})}\BibitemShut {NoStop}%
\bibitem [{\citenamefont {Liang}\ \emph {et~al.}(2015)\citenamefont {Liang},
  \citenamefont {Xiao}, \citenamefont {Li}, \citenamefont {Zhu}, \citenamefont
  {Zhu}, \citenamefont {Bao}, \citenamefont {Zhou},\ and\ \citenamefont
  {Wu}}]{Felarge}%
  \BibitemOpen
  \bibfield  {author} {\bibinfo {author} {\bibfnamefont {J.}~\bibnamefont
  {Liang}}, \bibinfo {author} {\bibfnamefont {X.}~\bibnamefont {Xiao}},
  \bibinfo {author} {\bibfnamefont {J.}~\bibnamefont {Li}}, \bibinfo {author}
  {\bibfnamefont {B.}~\bibnamefont {Zhu}}, \bibinfo {author} {\bibfnamefont
  {J.}~\bibnamefont {Zhu}}, \bibinfo {author} {\bibfnamefont {H.}~\bibnamefont
  {Bao}}, \bibinfo {author} {\bibfnamefont {L.}~\bibnamefont {Zhou}},\ and\
  \bibinfo {author} {\bibfnamefont {Y.}~\bibnamefont {Wu}},\ }\bibfield
  {title} {\bibinfo {title} {Quantitative study of the quadratic
  magneto-optical kerr effects in fe films},\ }\href@noop {} {\bibfield
  {journal} {\bibinfo  {journal} {Optics express}\ }\textbf {\bibinfo {volume}
  {23}},\ \bibinfo {pages} {11357} (\bibinfo {year} {2015})}\BibitemShut
  {NoStop}%
\bibitem [{\citenamefont {Hamrlov{\'a}}\ \emph {et~al.}(2016)\citenamefont
  {Hamrlov{\'a}}, \citenamefont {Legut}, \citenamefont {Veis}, \citenamefont
  {Pi{\v{s}}tora},\ and\ \citenamefont {Hamrle}}]{FeCuNiQMOKE}%
  \BibitemOpen
  \bibfield  {author} {\bibinfo {author} {\bibfnamefont {J.}~\bibnamefont
  {Hamrlov{\'a}}}, \bibinfo {author} {\bibfnamefont {D.}~\bibnamefont {Legut}},
  \bibinfo {author} {\bibfnamefont {M.}~\bibnamefont {Veis}}, \bibinfo {author}
  {\bibfnamefont {J.}~\bibnamefont {Pi{\v{s}}tora}},\ and\ \bibinfo {author}
  {\bibfnamefont {J.}~\bibnamefont {Hamrle}},\ }\bibfield  {title} {\bibinfo
  {title} {Principal spectra describing magnetooptic permittivity tensor in
  cubic crystals},\ }\href@noop {} {\bibfield  {journal} {\bibinfo  {journal}
  {Journal of Magnetism and Magnetic Materials}\ }\textbf {\bibinfo {volume}
  {420}},\ \bibinfo {pages} {143} (\bibinfo {year} {2016})}\BibitemShut
  {NoStop}%
\bibitem [{\citenamefont {Jahn}(1973)}]{jahn1973linear}%
  \BibitemOpen
  \bibfield  {author} {\bibinfo {author} {\bibfnamefont {I.}~\bibnamefont
  {Jahn}},\ }\bibfield  {title} {\bibinfo {title} {Linear magnetic
  birefringence in the antiferromagnetic iron group difluorides},\ }\href@noop
  {} {\bibfield  {journal} {\bibinfo  {journal} {physica status solidi (b)}\
  }\textbf {\bibinfo {volume} {57}},\ \bibinfo {pages} {681} (\bibinfo {year}
  {1973})}\BibitemShut {NoStop}%
\end{thebibliography}%

\end{document}